\documentclass[12pt]{article}

\usepackage{placeins}
\usepackage{pifont}
\usepackage{graphicx}
\usepackage{booktabs}
\usepackage{multirow}
\usepackage{amsmath}
\usepackage{amssymb}
\usepackage{lmodern}
\usepackage{rotating}
\usepackage{float}
\usepackage{lscape}
\usepackage[T1]{fontenc}
\usepackage[a4paper,left=2.5cm,right=2.5cm,top=3cm,bottom=3cm]{geometry}
\usepackage[flushleft]{threeparttable}

\usepackage[style=authoryear-comp, backend=bibtex, maxcitenames=2, maxbibnames=99 ,dashed=false]{biblatex}
\bibliography{mediation_refs}

\usepackage{hyperref}

\begin{document}

\title{Causal Mediation Analysis with a Time-Dependent Mediator, Time-Dependent Confounders and a Time-to-Event Outcome: Revisiting the Difference Method}
\date{}
\author{Robin Denz and Nina Timmesfeld \\ \\ Ruhr-University Bochum \\ Department of Medical Informatics, Biometry and Epidemiology}

\maketitle

\begin{abstract}
	Mediation analysis is a powerful tool to decompose treatment effects into direct and indirect components, enabling explanations of total treatment effects in a formal statistical framework. However, applying such analyses to settings with time-to-event outcomes, time-dependent mediators and confounders remains challenging. Existing methods are statistically complex, computationally intensive, and rarely available in user-friendly software. The difference method offers a simple alternative, but its performance in this setting has not been systematically evaluated. We conducted a simulation study and real-world data analysis to fill this gap in the literature. Using Cox proportional hazards, Aalen additive hazards, and accelerated failure time (AFT) models with time-varying covariates, the difference method was compared across different data generation processes with time-dependent mediators and confounders, focusing on bias in estimated indirect effects. The parametric mediational g-formula was used as benchmark comparator. If correctly specified, the Aalen model based difference method produced unbiased estimates in the absence of a time-dependent confounder that was directly caused by the treatment. Similar results were obtained when using the Cox model based difference with rare outcomes, but not with common outcomes. The AFT model based difference method was biased in almost all scenarios, due to collapsibility issues. Only the parametric mediational g-formula was unbiased in all scenarios. In contrast to specialized methods, the difference method requires additional, often unrealistic, assumptions, such as the absence of a direct causal relationship of the treatment on time-dependent confounders. If those assumptions hold, however, it may be used as a simple and efficient alternative.
\end{abstract}

\section{Introduction} \label{sec::introduction}

Mediation analysis is a central tool to investigate the mechanisms by which some treatment causally influences an outcome. Its main goal is to decompose the total effect of the treatment into \emph{direct} and \emph{indirect} components, with respect to a mediating variable, which is suspected to be caused by the treatment and may in turn influence the outcome \parencite{VanderWeele2015}. These decompositions are valuable for designing interventions and deepening scientific understanding beyond total treatment effects. As such, methods for mediation analysis are increasingly being used and developed \parencite{Rijnhart2021, LapointeShaw2018}. A comprehensive overview is given by VanderWeele \parencite{VanderWeele2015}.
\par
Originally, these methods were developed only for continuous mediators and outcomes measured at single points in time, but have been extended to cover many other settings since \parencite{VanderWeele2015, Baron1986}. Extensions for \emph{time-to-event} outcomes, such as the time until death or disease progression, are particularly important for medical research. These types of outcomes are often \emph{right-censored}, meaning that for some individuals it is only known that the event of interest has not occurred up until some point in time, which considerably complicates mediation analysis \parencite{LapointeShaw2018, Lange2011, Fulcher2017}. Additionally, the longitudinal nature of such time-to-event data almost inevitably brings variables that vary over time. For example, the effect of Ursodeoxycholic acid on survival of patients with primary biliary cholangits might be mediated by histological progression or other time-dependent markers of disease worsening \parencite{Rudic2012}. When mediators and confounders are time-dependent, both the definition of appropriate target estimands and their estimation becomes particularly challenging \parencite{Didelez2019, Mogensen2023, Aalen2018}.
\par
Multiple methods to perform causal mediation analysis with a time-dependent mediator and a time-to-event outcome have been proposed in the literature \parencite{Kormaksson2024, Lin2017, Lin2017a, Vansteelandt2019, Kateline2025, Aalen2020, Breum2024, Huang2021, Gao2023, Valeri2023, Bhandari2025, Bhandari2025a, Zeng2023, Wang2025}. Some are based on various g-methods, such as the parametric mediational g-formula, and / or dynamic path analysis, \parencite{Kormaksson2024, Lin2017, Lin2017a, Vansteelandt2019, Kateline2025, Aalen2020, Fosen2006} while others use the semi-competing risk or multi-state modelling framework, \parencite{Breum2024, Huang2021, Gao2023, Valeri2023} complex Bayesian joint models based on Markov chain Monte-Carlo simulation \parencite{Bhandari2025, Bhandari2025a, Zeng2023} or targeted maximum likelihood estimation \parencite{Wang2025}.
\par
Although these methods offer valuable estimation strategies, it is not straightforward to apply them in practice. First, they all require advanced knowledge of both statistical and causal inference based concepts. Secondly, most of these methods are not implemented in user-friendly software packages. Finally, many of the proposed methods also require models to be fit sequentially for each relevant point in time, sometimes for all involved variables, which is computationally demanding \parencite{Vansteelandt2019, Aalen2020}. Taken together with the need for non-parametric bootstrapping to obtain valid confidence intervals and standard errors, \parencite{Efron1979} they often require long computation times. If hundreds or thousand of points in time need to be considered, their usage may even be infeasible in practice.
\par
It may thus be tempting to use the \emph{difference method} instead, which is a simple and widely known approach that has been used extensively in settings without time-varying mediators \parencite{VanderWeele2015, Susser1973, Jiang2015}. Briefly, this method utilizes two models for the outcome: one for the direct effect (including both the treatment, mediator and confounders as independent variables) and one for the total effect (including only the treatment and confounders as independent variable). The difference of the obtained beta coefficients is used to estimate the indirect effect \parencite{VanderWeele2015}. Extending this approach to time-to-event outcomes with time-varying mediators and confounders is straightforward from a computational perspective \parencite{Vansteelandt2019}. For example, users may simply fit two Cox proportional hazards models~\parencite{Cox1972} with time-dependent covariates, which requires only two lines of code with standard software \parencite{Therneau2025}.
\par
The appeal of such simplicity is clear. Nevertheless, the usage of this method is controversial for multiple reasons. First, studies on time-to-event outcomes with time-fixed mediators have shown that, depending on the underlying data generation process (DGP) and the employed models, the difference method may produce biased estimates \parencite{Lange2011, Fulcher2017}. For example, \textcite{Lange2011} showed that the Cox proportional hazard model based difference method does not work well with common outcomes. Furthermore, it is well understood that under certain types of time-dependent feedback between the treatment, mediator, and confounders, the estimation of marginal causal effects using conditional models is not possible, complicating the use of the difference method for mediation analysis \parencite{Lin2017a, Vansteelandt2019, Hernan2020, Daniel2011}. As a result, the method is generally discouraged in this context. However, there has been no systematic investigation of its empirical performance in realistic data-generating scenarios with time-dependent mediators for a well-defined target estimand. Beyond the known bias under violations of structural assumptions, the properties of the method in this context are thus unclear. This paper seeks to fill that gap.
\par
Through a simulation study and the analysis of data from a randomized controlled trial on the effect of Ursodeoxycholic acid on survival in primary biliary cholangitis patients, we evaluate the performance of the difference method for causal mediation analysis in the presence of time-dependent mediators, time-dependent confounders, and a time-to-event outcome. We apply the difference method using Cox proportional hazards models, \parencite{Cox1972} Aalen additive hazards models \parencite{Aalen1989} and accelerated failure time (AFT) models, \parencite{Wei1992} comparing the results to the known truth and against the parametric mediational g-formula, \parencite{Lin2017a} which serves as a benchmark for more rigorous methods. Our aim is to identify conditions under which the difference method produces unbiased estimates of direct and indirect effects, and to highlight scenarios where it yields severely biased results. In doing so, we aim to provide applied researchers with clearer guidance on the potential risks and limitations of relying on this deceptively simple method in complex longitudinal survival settings.

\section{Motivating Example} \label{sec::example_introduction}

Before diving into the methodological details, we first want to introduce a scenario, in which the sort of mediation analysis discussed in this paper may be of interest to practitioners. Roughly, this application focuses on disentangling direct and indirect effects of a medical treatment in patients with primary biliary cholangitis (PBC), which is a rare immune-mediated chronic cholestatic liver disease \parencite{Hohenester2009}. It is characterized by the progressive destruction of small intrahepatic bile ducts, leading to impaired bile flow, chronic cholestasis, and ultimately liver fibrosis and cirrhosis if left untreated \parencite{Hohenester2009, Trivella2023}.
\par
The primary treatment for PBC is Ursodeoxycholic acid (UDCA), which aims to slow disease progression and improve long-term outcomes. Despite some initial controversy fuelled by early meta-analyses,\parencite{Goulis1999, Rudic2012} UDCA has been shown to significantly increase liver transplant-free survival (LTFS) time in multiple studies \parencite{Trivella2023, Lammers2014, Harms2019, Zhu2015a}. For example, in an international cohort study including 3902 PBC patients from eight countries in Europe and North America, an adjusted hazard ratio of 0.46 (95\% CI 0.40--0.52) was observed when comparing LTFS among UDCA-treated and untreated patients \parencite{Harms2019}. Although the benefits of UDCA for PBC patients are well documented, the precise mechanism by which UDCA improves these outcomes is not as well understood \parencite{Hohenester2009, Beuers2006, Paumgartner2004}. A fairly consistent finding in the literature is that UDCA slows histological progression, \parencite{Rudic2012} which likely mediates the effect of UDCA on LTFS, but no formal causal mediation analysis has been conducted so far. To get a better understanding about the UDCA treatment, it would be of great benefit to quantify the direct and indirect effects of treatment with UDCA on LTFS with respect to histological progression by two stages in PBC patients.
\par
In this scenario, UDCA is the baseline treatment, histological progression is the time-dependent mediator and the LTFS is the time-to-event outcome. We performed such an analysis to illustrate the potential usage and pitfalls of the difference method and the parametric mediational g-formula in a setting with a time-to-event outcome and time-dependent mediators and confounders. We used publicly available data from 170 patients that were enrolled in a double-blind, placebo-controlled trial of UDCA at the Mayo Clinic from 1988 to 1992, conducted by \textcite{Lindor1994}. The data is freely available as part of the \texttt{survival} R package, with ten patients from the original study being excluded due to incomplete follow-up \parencite{Therneau2000}. A detailed discussion of the study results are given in the primary article \parencite{Lindor1994}. The aim of this analysis is not to generate new clinical evidence. In fact, the sample size is not sufficient for reliable results to be drawn and some required identifiability assumptions likely do not hold for either analysis, as discussed below. Instead, the presented analysis is merely meant as an illustrative example to showcase how the investigated methods may be used in practice. More details and results of this analysis are discussed in section~\ref{sec::example}.

\section{Background} 

\subsection{Notation and setup}

\begin{figure}[!htb]
	\centering
	\includegraphics[width=0.6\linewidth]{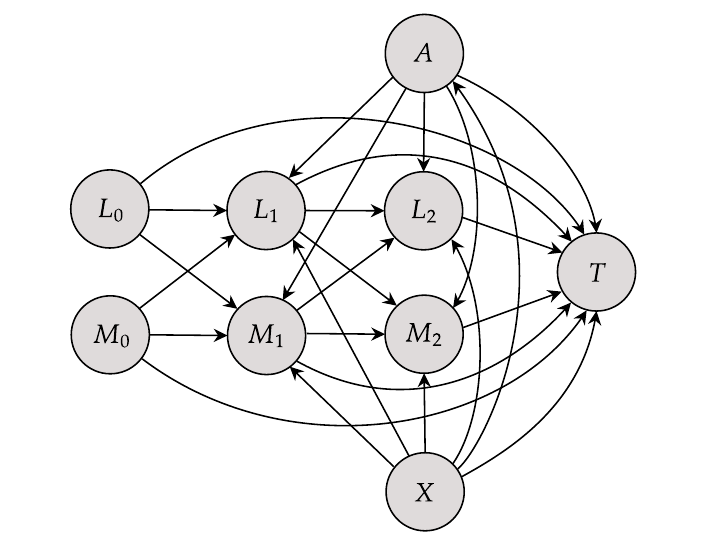}
	\caption{A causal directed acyclic graph describing the assumed data generation process for the baseline treatment $A$, baseline confounder $X$, the time-dependent variable $L_{t}$, the time-dependent mediator $M_t$ and the time-to-event outcome $T$. The graph is only drawn for $t \in \{0, 1, 2\}$, showing only one time-dependent confounder and omitting the independent error terms, for simplicity. Here, $L_t$ acts as a time-dependent confounder for the relationship between $M_t$ and $Y_{t+1}$.}
	\label{fig::dag_dgp}
\end{figure}

Let $A$ be a binary baseline treatment variable with $A \in \{0, 1\}$, where 0 indicates the control condition and 1 the treatment condition. Similarly, let $M_t$ be a time-dependent mediator and $Y_t$ be a binary time-dependent outcome of interest, with $t$ specifying the respective point in time. Additionally, let $X$ denote a baseline covariate and let $\mathbf{L}_{t} = \{L_{1,t}, L_{2,t}, L_{i,t}, ..., L_{w,t}\}$ denote a vector of $w$ time-dependent variables (indexed by $i$), that may act as confounders of the causal relationship between $M_t$ and $Y_{t+1}$, as shown in figure~\ref{fig::dag_dgp}. To keep the notation simple, we focus on a single baseline variable and equi-distant discrete-time measurements of $\mathbf{L}_{t}$, $M_t$ and $Y_t$, with $t \in \{0, 1, 2, ..., k\}$ for every individual.
\par
The first time at which $Y_t$ turns one is denoted as $T$. In practice, this time might not be observed for all individuals, due to loss to follow-up or other censoring events that may occur before $T$. Let the time until such censoring events be denoted $C$. In practice, we only observe $T^{obs} = \min(T, C)$, with corresponding event indicator $D^{obs} = I(T \leq C)$, where $I()$ denotes the indicator function. The data one would observe under this setup, are thus equivalent to data from a cohort study or data from secondary databases, such as administrative health-claims data.
\par
We assume that the DGP can be described by a non-parametric structural equation model with independent error terms, \parencite{Pearl2009} with all temporal orderings being defined according to the causal directed acyclic graph (DAG) in figure~\ref{fig::dag_dgp}. For simplicity, we assume a Markovian property, so that, given all variables at $t-1$, variables at $t$ are independent of the past. In the subsequent simulation study, this DAG is changed slightly, to allow investigation of different scenarios, but the general temporal orderings are consistent throughout the article. We additionally assume independent censoring of $T^{obs}$ given time-independent confounders $X$, e.g. $T \perp C | X$.

\subsection{Target estimand} \label{sec::target_estimand}

Estimands for mediation analysis are usually defined in terms of contrasts of \emph{counterfactual} or \emph{potential} outcomes, denoting the outcome one would observe under hypothetical interventions on some variables \parencite{Pearl2009, Hernan2016}. In standard mediation analysis without time-dependent mediators and confounders, the classic definition of \emph{natural effects}, as proposed by \textcite{Robins1992} relies on the nested counterfactual $Y^{a,M^{a'}}$, which denotes the outcome that would be observed if the treatment was set to $a$, but the mediator was set to the value it would have taken under $a'$ \parencite{Didelez2019, Robins1992}. By contrasting $Y^{a,M^{a'}}$ under different value combinations of $a$ and $a'$, direct and indirect effects may be defined \parencite{VanderWeele2015}.
\par
Unfortunately, this strategy to define the estimand does not generalize to the setting considered in this article. First, it is possible that a patient survives longer under treatment $a$ than he/she would have under treatment $a'$, making the mediator process in one of these counterfactual worlds ill-defined \parencite{Didelez2019}. Additionally, because the observation of the time-dependent confounders $\mathbf{L}_{t}$ and mediator $M_t$ depends on prior survival, they act as post-treatment confounders, which makes the identifiability of natural effects impossible \parencite{Avin2005}.
\par
To circumvent these issues, we follow the strategy of \textcite{Vansteelandt2019} to define the target estimand as \emph{path-specific effects}, \parencite{Avin2005, Pearl2001} by first recursively defining counterfactual values for $\mathbf{L}_{t}$, $M_t$ and $Y_t$, after intervention on $A$. Let $\mathcal{L}_{i,t}(a, l_{i, t-1}, m_{t-1}, u_{i,t})$ denote a function to evaluate the counterfactual value of $L_{i,t}$, given that $A$ had been set to $a$, $L_{i, t-1}$ had been set to $l_{i, t-1}$ and $M_{t-1}$ had been set to $m_{t-1}$, with the exogenous error term taking value $u_{i,t}$. Here, $U_{i,t}$ captures all unobserved sources of variation in the system and shared across counterfactual worlds. Let $\mathcal{M}_t(a, \mathbf{l}_{t-1}, m_{t-1}, u_{M,t})$ and $\mathcal{Y}_t(a, \mathbf{l}_{t-1}, m_{t-1}, y_{t-1}, u_{Y,t})$ be defined accordingly. The counterfactual $L_{i,t}^{a, a'}$ is then defined as:

\begin{equation}
	L_{i,t}^{a, a'} = \begin{cases}
		L_{i,0}, & \text{if } t = 0 \\
		\mathcal{L}_{i,t}\!\left(a, L_{i, t-1}^{a, a'}, M_{t-1}^{a, a'}, U_{i,t}\right) & \text{if } t > 0 \\
	\end{cases}, \label{eq::L_a_a'}
\end{equation}

with:

\begin{equation}
	M_t^{a, a'} = \begin{cases}
		M_0 & \text{if } t = 0 \\
		\mathcal{M}_t\!\left(a', \mathbf{L}_{t-1}^{a, a'}, M_{t-1}^{a, a'}, U_{M,t}\right) & \text{if } t > 0
	\end{cases}. \label{eq::M_a_a'}
\end{equation}

In words, $L_{i,t}^{a, a'}$ is defined as the value of $L_{i,t}$ that would have been observed if the time-dependent confounder process $L_{i,t}$ had evolved as if $A$ had been set to $a$, but the mediator process had evolved as if $A$ had been set to $a'$, both using the updated values of each other in a feedback over time. For example, given these equations with $w = 1$, the counterfactual value of $L_{1,2}^{a, a'}$ is defined as:

\begin{equation}
	L_{1,2}^{a, a'} = \mathcal{L}_{1,2}\!\left(a, \mathcal{L}_{1,1}\!\left(a, L_{1,0}, M_0, U_{1,1}\right), \mathcal{M}_1\!\left(a', L_{1,0}, M_0, U_{M,1}\right), U_{1, 2}\right).
\end{equation}

This is the value of $L_{1,2}$ that would be observed if $A$ had been set to $a$, with $M_1$ taking the value one would see if $A$ had been set to $a'$ instead, given the observed values of $L_{1,0}$ and $M_0$, and with $L_{1, 1}$ taking its counterfactual value under $a$, given $L_{1,0}$ and $M_0$ . The counterfactual outcome $Y_t^{a, a'}$ is defined accordingly as:

\begin{equation}
	Y_t^{a, a'} = \begin{cases}
		Y_0 & \text{if } t = 0 \\
		\mathcal{Y}_t\!\left(a, \mathbf{L}_{t-1}^{a, a'}, M_{t-1}^{a, a'}, Y_{t-1}^{a, a'}, U_{Y,t}\right) & \text{if } t > 0
	\end{cases}, \label{eq::Y_a_a'}
\end{equation}

with a similar interpretation. Using this definition, the potential outcomes $Y_t^{0, 0}$ and $Y_t^{1, 1}$ define the values of $Y_t$ that would be observed at $t$ if $A$ was set to 0 or 1 respectively, leaving everything else unchanged. The outcome $Y_t^{1, 0}$ on the other hand, denotes the value of $Y_t$ that would be observed at $t$ if the treatment had been set to $A = 1$, but the mediator process had evolved as if $A$ had been set to 0, leaving the process of the time-dependent confounders $L_{i,t}$ unchanged, with the exception of it being dependent on the changed mediator process instead of the mediator process under $A = 0$. In all of these counterfactuals, censoring events are not directly removed or intervened on \parencite{Wen2025}.
\par
If the time-dependent confounders include the at-risk indicator $I(T > t)$, the issue of post-treatment confounding is resolved, because the mediator under $a'$ is set to the level it would have taken if the time-dependent confounders, including the survival status, had evolved as if $a$. A more thorough discussion is given by \textcite{Vansteelandt2019} Using very small steps in time, these discrete-time based definitions approximate their continuous time analogues \parencite{Kateline2025}.
\par
On the basis of these potential outcomes, the \emph{total effect} (TE), \emph{path-specific direct effect} (DE) and \emph{path-specific indirect effect} (IE) can now be defined on different scales. Let $T^{a, a'}$ denote the first time at which $Y_t^{a, a'}$ turns one. The hazard function of $T^{a, a'}$ is then given as:

\begin{equation}
	\lambda^{a,a'}(t) = P\left(T^{a,a'} = t \,\middle|\, T^{a,a'} \ge t \right).
\end{equation}

Using this definition, the target estimands on the log hazard ratio scale are defined as:

\begin{align}
	\Psi_{TE}(t) = & \log\!\left(\frac{\lambda^{1, 1}(t)}{\lambda^{0, 0}(t)}\right), \label{eq::target_TE} \\
	\Psi_{DE}(t) = & \log\!\left(\frac{\lambda^{1, 0}(t)}{\lambda^{0, 0}(t)}\right), \label{eq::target_DE} \\
	\Psi_{IE}(t) = & \log\!\left(\frac{\lambda^{1, 1}(t)}{\lambda^{1, 0}(t)}\right). \label{eq::target_IE}
\end{align}

Assuming a constant hazard ratio, as is usually done in a Cox model, the time-dependency could simply be dropped. In this study, we additionally focus on the hazard difference scale and the log survival time scale (see appendix for formal definitions). These three scales were chosen, because they correspond to the scale of the beta coefficients for popular time-to-event models \parencite{Cox1972, Aalen1989, Wei1992}. Alternatively, the difference in counterfactual survival probabilities, \parencite{Vansteelandt2019, Denz2023} or difference in restricted mean survival times~\parencite{Chernofsky2025} could be used as well. Because the target estimands are defined as expected outcomes over a population without regards to the observed values of $A$, these effects may be considered marginal average treatment effects \parencite{Kateline2025}.
\par
As explained by \textcite{Vansteelandt2019}, effects defined according to this strategy make no mention of the specific values of mediators at single points in time. Instead, the effect of interest is the effect mediated through the entire mediator process up to $t$. Note also that, the indirect effect defined according to equation \ref{eq::target_IE} deliberately does not include indirect effects that occur due to $A$ influencing $\mathbf{L}_{t}$, which then influences $M_{t+1}$, which in turn influences $Y_{t+2}$. Alternative path-specific effects are given by \textcite{Kateline2025} Additionally, other authors proposed different definition strategies, such as decomposing the treatment into two separate different components which can be intervened on separately \parencite{Didelez2019, Hernan2020} or principal stratum based effects \parencite{Gao2023}. For simplicity, we exclusively focus on the ones derived from the potential path-specific outcomes~\ref{eq::L_a_a'}, \ref{eq::M_a_a'} and \ref{eq::Y_a_a'} in this study, but others might be more appropriate depending on the setting or study question.

\subsection{Identifiability assumptions} \label{sec::assumptions}

Because the target estimands are defined in terms of contrasts in potential outcomes, they are not directly observable in reality. To allow identification ot these estimands from the actually observed data, multiple identifiability assumptions have to be made. In particular, the \emph{counterfactual consistency}, \parencite{Cole2009, VanderWeele2009} \emph{positivity}, \parencite{Westreich2010} \emph{sequential ignorability} \parencite{Deng2024a, Forastiere2018} and \emph{cross-world independence} \parencite{Andrews2021} assumptions are required. These are described and discussed in great detail elsewhere \parencite{Mogensen2023, Kateline2025, Nguyen2022}.
\par
Briefly, the counterfactual consistency assumption states that the observed value of the outcome under treatment $a$ is equal to the potential outcome that would have been observed under treatment $a$ \parencite{Cole2009, VanderWeele2009}. Further, the positivity assumption states, that each individual must have a positive probability below 1 to have any possible time-dependent confounder or mediator history. The sequential ignorability assumption is concerned with different kinds of confounding in the data generation process. It states that, conditional on observed time-fixed and time-dependent confounders (i) there is no confounding of the $A$ and $Y_{t+1}$ relationship, (ii) there is no confounding of the $A$ and $M_t$ relationship, (iii) there is no confounding of the $M_t$ and $Y_{t+1}$ relationship, and (iv) there is no confounding of the $\mathbf{L}_t$ and $M_{t+1}$, or $M_{t}$ and $\mathbf{L}_{t+1}$ relationships \parencite{Kateline2025}. Finally, the cross-world independence assumption states that the potential outcomes under intervention on both $A$ and $M_t$ is independent of the value of $M_t$ after intervention on $A$ \parencite{Hernan2020, Deng2024a, Nguyen2022}.
\par
Further model- and method-specific assumptions will be necessary for unbiased effect estimation, as discussed below.  

\section{Methods} \label{sec::methods}

This study focuses on the \emph{difference method} \parencite{VanderWeele2015, Susser1973} and on the \emph{parametric mediational g-formula} \parencite{Lin2017a, Keil2014}. The former is the main point of interest of this study and the latter was chosen as reference, representing more advanced alternatives. For the parametric mediational g-formula, we also use the terms \emph{parametric g-formula} or just \emph{g-formula} interchangeably throughout this article.

\subsection{The difference method}

The difference method works by fitting two models: one for the direct effect and one for the total effect of $A$ on $T$. In the considered setting, the direct effect model contains the treatment $A$, both time-fixed and time-dependent confounders such as $X$ and $\mathbf{L}_{t}$, and the time-dependent mediator $M_t$ as time-varying independent variables, so that the updated values are used at the correct points in time \parencite{Zhang2018}. The total effects model should contain only the treatment $A$ and baseline confounders such as $X$. The resulting (possibly time-dependent) beta coefficients for $A$ can then serve as estimates of the direct and total effects, assuming no interaction between $A$ and $M_t$. The indirect effect is obtained by simply subtracting the direct effect from the total effect \parencite{VanderWeele2015, Susser1973}. Formally:

\begin{align}
	\hat{\Psi}_{TE}(t) = & \enspace \hat{\theta}_{A}(t), \\
	\hat{\Psi}_{DE}(t) = & \enspace \hat{\beta}_{A}(t), \\
	\hat{\Psi}_{IE}(t) = & \enspace \hat{\theta}_{A}(t) -  \hat{\beta}_{A}(t),
\end{align}

where $\hat{\theta}_{A}(t)$ is the estimated coefficient for $A$ at $t$ in the total effects model and $\hat{\beta}_{A}(t)$ is the estimated coefficient for $A$ at $t$ in the direct effects model. If the coefficients are not time-dependent, the corresponding time dependency is simply dropped. Different types of models may be used for their estimation. Popular alternatives are the Cox proportional hazards model, \parencite{Cox1972} the Aalen additive hazards model \parencite{Aalen1989} and AFT models \parencite{Wei1992}. Depending on the model, estimates will be on different scales. When using Cox models of the form:

\begin{equation}
	\lambda_{DE}(t) = \lambda_0(t) \exp\!\big( \beta_{A}A + \beta_{M}M_t + \beta_{L_1}L_{1,t} + \beta_{L_2}L_{2,t} + ... + \beta_{L_w}L_{w,t} + \beta_{X}X\big),
\end{equation}

and:

\begin{equation}
	\lambda_{TE}(t) = \lambda_0(t) \exp\!\big(\theta_{A}A + \theta_{X}X\big),
\end{equation}

the resulting estimates are on the log hazard ratio scale. Usually, time-constant coefficients are assumed, although this assumption may be relaxed if needed \parencite{Zhang2018}. Note that if the model for the direct effect is true and there is an indirect effect of $A$ on $T$ through $A \rightarrow M_t \rightarrow Y_{t+1}$, the coefficient $\theta_{A}$ in the total effects model is likely not actually constant, because of the potentially changing distribution of $M_t$ over time. $\theta_{A}$ should then be interpreted as the average log hazard ratio of $A$ in the observed follow-up time. In contrast, when using Aalen models of the form \parencite{Aalen1989}:

\begin{equation}
	\lambda_{DE}(t) = \lambda_0(t) + \beta_{A}(t)A + \beta_{M}(t)M_t + \beta_{L_1}(t)L_{1,t} + \beta_{L_2}(t)L_{2,t} + ... + \beta_{L_w}(t)L_{w,t} + \beta_{X}(t)X, 
\end{equation}

and:

\begin{equation}
	\lambda_{TE}(t) = \lambda_0(t) + \theta_{A}(t)A + \theta_{X}(t)X,
\end{equation}

the results would be directly on the hazard difference scale. Contrary to the Cox model time-dependent coefficients are often used here, as shown in the model equations. It is, however, also possible to fit such models with constant beta coefficients \parencite{Aalen1989}. Finally, the AFT models have the form \parencite{Wei1992}:

\begin{equation}
	T_{DE} = \exp\!\big( \beta_0 + \beta_{A}A + \beta_{M}M_t + \beta_{L_1}L_{1,t} + \beta_{L_2}L_{2,t} + ... + \beta_{L_w}L_{w,t} + \beta_{X}X\big) \cdot \exp(\sigma \varepsilon), 
\end{equation}

and:

\begin{equation}
	T_{TE} = \exp\!\big(\theta_0 + \theta_{A}A + \theta_{X}X\big) \cdot \exp(\sigma \varepsilon),
\end{equation}

with $\beta_0$ and $\theta_0$ being the intercept, $\sigma$ being a scaling factor and $\varepsilon$ being a random error term that follows some parametric distribution. In contrast to the Cox and Aalen models, the AFT models directly model the survival time instead of the hazard. The coefficients are thus on the log survival time scale, making them interpretable as accelerating (or decelerating) factors \parencite{Wei1992}. As in the Cox model, if the direct effects model is true and indirect effects through  $A \rightarrow M_t \rightarrow Y_{t+1}$ exist, $\theta_{A}$ should usually be interpreted as an average over the observed follow-up time, if not directly modelled as time-dependent coefficient.

\subsection{Issues of the difference method}

Although the difference method is very popular in epidemiology and other fields when considering continuous outcomes and time-fixed mediators, \parencite{VanderWeele2015} its usage with time-to-event outcomes or time-dependent mediators is more controversial \parencite{Lange2011, Fulcher2017, VanderWeele2011}. When using Cox models, multiple potential issues arise. First, Cox models are generally non-collapsible, in the sense that the inclusion or exclusion of an independent variable in the model may change the value of the coefficients, even if that variable is not a confounder \parencite{Daniel2020, Martinussen2013}. Observed differences between $\hat{\beta}_{A}$ and $\hat{\theta}_{A}$ may therefore not accurately reflect indirect effects. In contrast, the coefficients of the Aalen and AFT models have been shown to be collapsible in previous studies dealing with time-fixed mediators \parencite{VanderWeele2015, Martinussen2013, Crowther2023, Ochoa2020}.
\par
Similarly, unobserved heterogeneity, e.g. omitting independent variables from the model that cause the outcome, may further bias results when using Cox models, if the true treatment effect of $A$ is not 0, due to built-in selection bias through conditioning on survival up to $t$ \parencite{Hernan2010}. Recent studies indicate, however, that this bias is small in practice \parencite{Strobel2023, Abrahamowicz2025}. Additionally, previous studies have shown that the Cox model based difference method with time-fixed mediators only produces unbiased estimates when the outcome is rare, limiting its usage with common outcomes. Neither the Aalen nor the AFT model based difference method required this rare outcome assumption with time-fixed mediators \parencite{VanderWeele2011, Tein2003}. However, when using AFT models, biased estimates may be produced by model misspecification arising from censoring or truncation \parencite{Fulcher2017, Ochoa2020}.
\par
Regardless of the form of the model used, some structural aspects of the underlying DGP may also prohibit the use of the difference method. As has been pointed out by other authors, \parencite{Daniel2011} if the time-dependent variables $\mathbf{L}_{t}$ are confounders of the relationship between $M_t$ and $Y_{t+1}$, and are simultaneously caused by $A$, the difference method generally produces biased estimates. The reason is that while including $\mathbf{L}_{t}$ as independent variables in the direct effects model removes the confounding effect of $\mathbf{L}_{t}$, it also removes parts of the actual direct effect. That is, it mistakenly adjusts for effects that flow through $A \rightarrow \mathbf{L}_{t} \rightarrow Y_{t+1}$, which should be counted as direct effects of $A$ on $Y_t$ in regards to $M_t$, because they are not mediated by $M_t$.
\par
The practical extent of these issues when applying the difference method with time-varying mediators is currently unknown. This study aims to quantify them through a comprehensive simulation study.

\subsection{The parametric mediational g-formula} \label{sec::methods_g_form}

A more rigorous method is the \emph{parametric mediational g-formula}, \parencite{Lin2017a} which is an extension of the regular g-formula for time-dependent exposures introduced by \textcite{Robins1986}. Let $\bar{\mathbf{L}}_{t}$ be defined as the history of $\mathbf{L}_{t}$ up to $t$ with $\bar{\mathbf{L}}_{t} = \left(\mathbf{L}_{0}, \mathbf{L}_{1}, ..., \mathbf{L}_{t}\right)$ and let $\bar{M}_t$ and $\bar{Y}_t$ be defined accordingly as $\bar{M}_t$ = ($M_0$, $M_1$, ..., $M_{t})$ and $\bar{Y}_t$ = ($Y_0$, $Y_1$, ..., $Y_{t})$, respectively. For simplicity, suppose that there is only one categorical baseline confounder $X$ and that both $\mathbf{L}_t$ and $M_t$ are binary variables. We further assume that the data are generated according to the DAG shown in figure~\ref{fig::dag_dgp} and that $Y_0 = 0$ for all individuals. The expectation of the counterfactual outcome $Y_t^{a, a'}$ can then be defined as \parencite{Lin2017a, Vansteelandt2019}:

\begin{equation}
	\begin{aligned}
		\mathbb{E}\!\left(Y_t^{a,a'}\right)
		&=
		\sum_{x} \sum_{\bar{\mathbf{l}}_{t-1}} \sum_{\bar{m}_{t-1}} \sum_{\bar{y}_{t-1}}
		P\left(Y_t = 1 \,\Big|\, a,\ \mathbf{l}_{t-1},\ m_{t-1},\ y_{t-1},\ x \right) \\
		&\quad \times \prod_{s=1}^{t-1}
		P\Big( \mathbf{l}_s \mid a,\ \mathbf{l}_{s-1},\ m_{s-1},\ x \Big) 
		\times \prod_{s=1}^{t-1}
		P\Big( m_s \mid a',\ \mathbf{l}_{s-1},\ m_{s-1},\ x \Big) \\
		&\quad \times \prod_{s=1}^{t-1}
		P\Big( y_s \mid a,\ \mathbf{l}_{s-1},\ m_{s-1}, y_{s-1}, \ x \Big)
		\times P(x)\; P(\mathbf{l}_0)\; P(m_0).
	\end{aligned}
\end{equation}

where, with a slight abuse of notation, we write $P(m | a', \bar{l}_{s-1}, \bar{m}_{s-1}, x)$ to denote $P(m | A = a', \bar{L}_{s-1} = \bar{l}_{s-1}, \bar{M}_{s-1} = \bar{m}_{s-1}, X = x)$ and use similar notation for all other conditional probabilities. The sums $\sum_{x}$, $\sum_{\bar l_{t}}$, $\sum_{\bar m_{t}}$ and $\sum_{\bar y_{t}}$ denote that the sum is to be taken over all possible values of $x$ and all possible $\bar{l}_t$-, $\bar{m}_t$ and $\bar{y}_t$-histories \parencite{Hernan2020}.
\par
In words, this equation states that the expectation of the counterfactual outcome $Y_t^{a, a'}$ can be written as a function of nested conditional probability densities. If either of $\mathbf{L}_t$, $M_t$ or $X$ is continuous, the corresponding sum may be replaced by an integral, substituting the probabilities for density functions. Theoretically, one thus only needs to estimate the conditional probabilities of $Y_t$, $L_{1,t}, L_{2,t}, ..., L_{w,t}$ and $M_t$, as well as the marginal distribution of $X$, $\mathbf{L}_0$ and $M_0$, to obtain an plug-in estimator of $\mathbb{E}(Y_t^{a, a'})$. Since neither of these quantities involve counterfactuals, they can be estimated from the observed data, under the stated identifiability assumptions. Using the same strategy, the target estimands given in section~\ref{sec::target_estimand} may be identified \parencite{Lin2017a, Vansteelandt2019}. Because it is difficult to calculate these values analytically, the estimation is usually performed using Monte-Carlo simulation methods \parencite{Lin2017a, Hernan2020, Daniel2011}.
\par
This works by first approximating the conditional probabilities using standard binomial regression models, in which the current value of the respective value is regressed against appropriately time-lagged values of itself and the relevant predictors. Depending on the parametric assumptions, this may be done with pooled models for all $t$ or with separate models for each $t$. These models are then used to directly simulate data under the respective counterfactual scenario, using the following steps:

\begin{enumerate}
	\item Set $t = 0$. For each individual $d = 1, ..., n$, set ($X_d, \mathbf{L}_{d, 0}, M_{d,0}, Y_{d,0}$) equal to their observed values.
	\item Increase $t$ by 1.
	\item Use the observed values of $X$ and the models for $\mathbf{L}_t$ to generate $n$ random values for $\mathbf{L}_t$ under $A = a$. Use the previously generated values for $M_{t-1}$ and $\mathbf{L}_{t-1}$ in the generation process.
	\item Use the observed values for $X$ and the model for $M_t$ to generate $n$ random values for $M_t$ under $A = a'$. Also use the previously generated values for $M_{t-1}$ and $\mathbf{L}_{t-1}$ in the generation process.
	\item Use the observed values for $X$ and the model for $Y_t$ to generate $n$ random values for $Y_t$ under $A = a$. Also use the previously generated values for $Y_{t-1}$, $M_{t-1}$ and $\mathbf{L}_{t-1}$ in the generation process.
	\item Repeat steps 2 to 5 until $Y_t$ is 1 for all $n$.
\end{enumerate} 

By repeating this procedure for different combinations of $a$ and $a'$, data under multiple counterfactual scenarios are simulated \parencite{Lin2017a, Vansteelandt2019}. Simple analysis methods may then be applied to these datasets to obtain estimates of the target estimands, depending on the desired scale. For example, if one wanted to estimate the indirect effect on the log hazard-ratio scale as defined in equation~\ref{eq::target_TE}, one would only need to (1) simulate data under $a = 1, a'= 0$ and $a = a' = 0$, (2) add an indicator for which scenario the dataset belongs to, (3) put the data together into one dataset and (4) fit a univariable Cox proportional hazards model, using the scenario indicator as the sole independent variable. Standard errors and confidence intervals may be obtained by bootstrapping this whole procedure (including the model fitting) \parencite{Keil2014}. Note that if the target estimand is defined without interventions on censoring events, censoring should also be added to the simulated data. Under the independent censoring assumption, this may be done by approximating an appropriate distribution of the censoring times using maximum likelihood estimation, sampling from this distribution and censoring the simulated event times appropriately.
\par
Although this method is conceptually simple and intuitive to understand, it is not trivial to apply in practice. First, it requires models for all of the involved variables, which may not be easy to obtain depending on the DGP and the available data. Secondly, the implementation of the approach does require software development knowledge or the use of specialized software, \parencite{Daniel2011} which may not be available. Finally, it is computationally expensive, especially when fitting models per point in time and when considering large amounts of points in time. If the Markovian assumption used here does not hold, the histories $\bar{\mathbf{L}}_t$, $\bar{M}_t$ and $\bar{Y}_t$ need to be used when fitting the models, further complicating estimation. Most of these hurdles can often be overcome in practice, but given the time and resource constraints faced by many applied researchers, simpler alternatives such as the difference method may seem more attractive.

\section{Simulation study} \label{sec::simulation}

\subsection{Data generation process} \label{sec::dgp}

Data for the simulation study was generated using a discrete-event simulation approach, based on a Gillespie type algorithm, \parencite{Gillespie1977, Anderson2007, Denz2025}. Six variables are included in the DGP, with three of those ($X$, $Z$, $A$) being time-fixed and the other three ($L_t$, $M_t$, $Y_t$) being binary and time-dependent. The former three consist of a standard normally distributed variable $X$, a Bernoulli distributed variable $Z$, and the binary treatment variable $A$, which is modelled as a log-binomial regression dependent on $Z$. The latter three are the time-dependent covariate $L_t$, the time-dependent mediator of interest $M_t$ and the time-dependent outcome of interest $Y_t$. Note that we use two specifications for $Y_t$: one that is consistent with both a Cox and AFT model and one that is consistent with an Aalen model, so that we can investigate the performance of each model without inherent model misspecification. Finally, the censoring time $C$ is generated using an exponential distribution. The DGP can be described using the following structural equations:

\begin{align}
	X \sim & \enspace N(0, 1), \label{eq::str_eq_A} \\[1ex]
	Z \sim & \enspace Bernoulli(0.5), \\[1ex]
	A \sim & \enspace Bernoulli\left(P_A(Z)\right), \\[1ex]
	L_t \sim & \enspace Bernoulli\!\left(P_L\!\left(t\,|\, A, X, L_{t-1}, M_{t-1}\right)\right), \\[1ex]
	M_t \sim & \enspace Bernoulli\!\left(P_M\!\left(t \,|\, A, X, Z, L_{t-1}, M_{t-1}\right)\right), \\[1ex]
	Y_t \sim & \enspace Bernoulli\!\left(P_Y\!\left(t \,|\, A, X, Z, L_{t-1}, M_{t-1}, Y_{t-1}\right)\right), \\[1ex]
	C \sim & \enspace \lceil Exp(\lambda_C) \rceil,
\end{align}

with the corresponding probability functions $P_A$, $P_L(t)$ and $P_M(t)$ being defined as:

\begin{align}
	P_A = & \enspace \exp\!\left( \beta_{0}^{A} + \beta_{Z}^{A}Z \right), \label{eq::P_A} \\[2ex]
	P_L(t) = &
	\begin{cases}
		0, & \text{if } t = 0 \\
		1, & \text{if } L_{t-1} = 1 \\
		\exp\!\left(
		\begin{aligned}
			& \log(0.01) + \beta_{A}^{L} A + \log(0.8) X + \\
			& \quad \beta_{M}^{L} M_{t-1}
		\end{aligned} \right) & \text{otherwise}\\
	\end{cases}, \label{eq::P_L} \\[2ex]
	\intertext{and:}
	P_M(t) = &
	\begin{cases}
		0, & \text{if } t = 0 \\
		1, & \text{if } M_{t-1} = 1 \\
		\exp\!\left(
		\begin{aligned}
			& \log(0.01) + \beta_{A}^{M} A + \log(0.8) X + \\
			& \quad \beta_{Z}^{M} Z + \beta_{L}^{M} L_{t-1} 
		\end{aligned} \right) & \text{otherwise}\\
	\end{cases}. \label{eq::P_M} \\[2ex]
\end{align}

Additionally, when investigating the performance of the Cox and AFT model based difference method, $P_Y(t)$ is defined as:

\begin{align}
	P_Y(t) = &
	\begin{cases}
		0, & \text{if } t = 0 \\
		1, & \text{if } Y_{t-1} = 1 \\		
		\exp\!\left(
		\begin{aligned}
			& \beta_{0}^{Y} + \log(0.9)A + \log(0.8)X + \\
			& \quad \beta_{Z}^{Y} Z + \log(2.5)L_{t-1} + \log(5)M_{t-1}
		\end{aligned} \right) & \text{otherwise}
	\end{cases}, \label{eq::P_Y_cox}
\end{align}

For the investigation of the Aalen model based difference method, it is instead defined as:

\begin{align}
	P_Y(t) = &
	\begin{cases}
		0, & \text{if } t = 0 \\
		1, & \text{if } Y_{t-1} = 1 \\		
		1 - \exp\!\left(-\!\left(
		\begin{aligned}
			& \beta_{0}^{Y} + \log(0.9)\gamma A + \log(0.8)\gamma X + \\
			& \quad \beta_{Z}^{Y}\gamma Z + \log(2.5)\gamma L_{t-1} + \log(5)\gamma M_{t-1}
		\end{aligned} \right)\right) & \text{otherwise}
	\end{cases}, \label{eq::P_Y_aalen}
\end{align}

where $\gamma = 0.00015$, representing a constant scaling factor, used to ensure that the relative strengths of the coefficients are comparable between the Cox / AFT model specification and the Aalen model specification. The exact values for the currently undefined beta coefficients, intercepts, and $\lambda_C$ depend on the considered simulation scenario, as discussed below. A table of all used values is given in the appendix. All of the resulting datasets have hundreds or thousand of distinct points in time for each individual, depending on the scenario and parameters. This mimics real datasets containing date-specific information, such as electronic health care records, patient registries or randomized controlled trials that track specific disease transitions.

\subsection{Scenarios} \label{sec::scenarios}

\begin{figure}[!htb]
	\centering
	\includegraphics[width=0.8\linewidth]{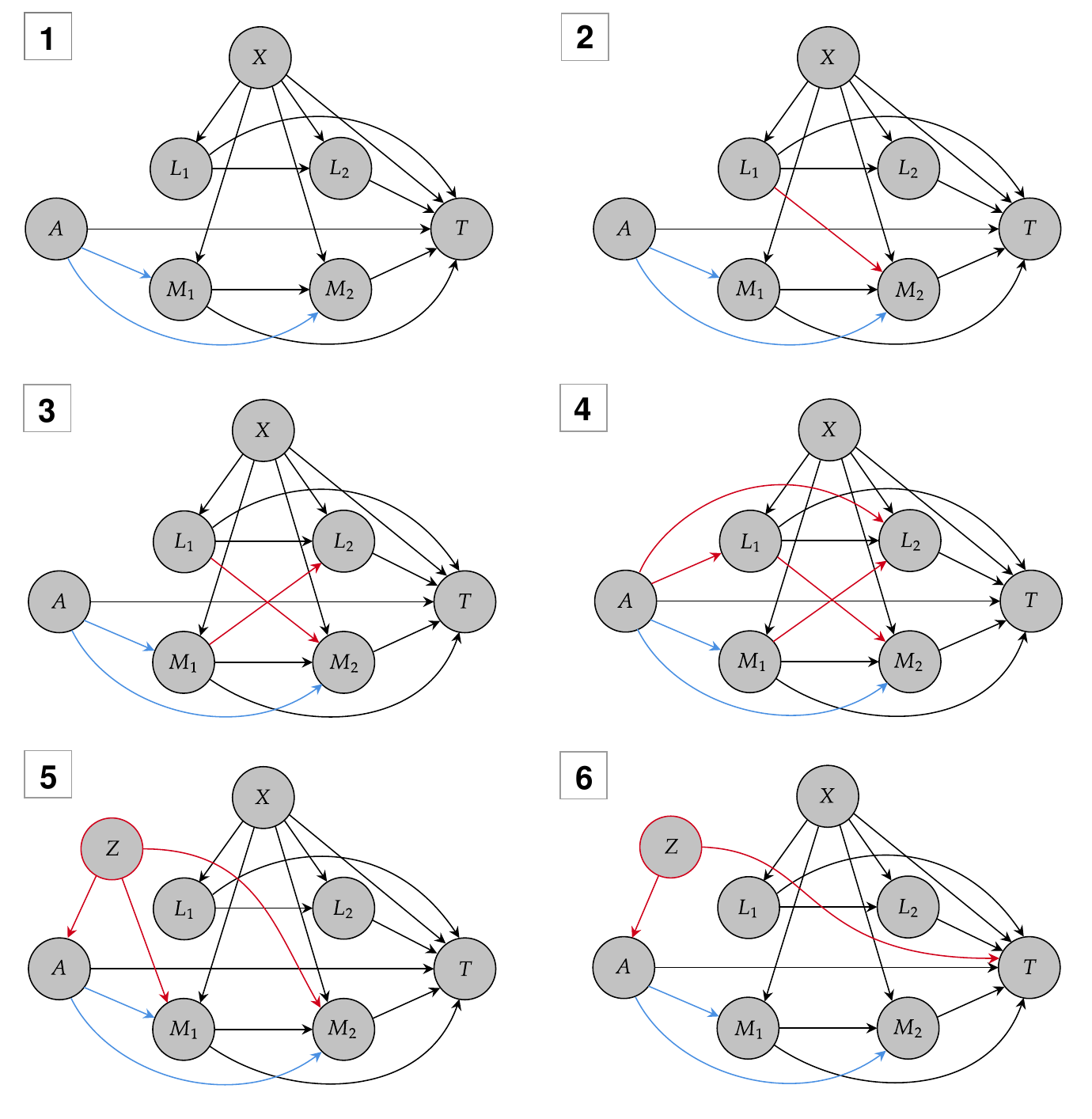}
	\caption{Causal directed acyclic graphs (DAGs) visualizing the causal relationships in the data generation processes used in each considered simulation scenario. All DAGs only show two points in time ($t = 1$ and $t = 2$) to make them more comprehensible. Values for $L_0$ and $M_0$ were omitted, because they are always 0 by definition. $A$ denotes the time-invariant treatment, $X$ and $Z$ denote time-invariant covariates, $L_t$ is a time-dependent covariate, $M_t$ is the time-dependent mediator and $T$ is the time until the first event occurring in the time-dependent outcome $Y_t$. Arrows coloured in blue are varied in all scenarios, while red arrows and outlines denote causal dependencies that are not present in scenario 1, to highlight differences between the scenarios.}
	\label{fig::dag_scenarios}
\end{figure}

Six data generation scenarios were considered, in which different types of confounding and time-dependent feedback loops between variables are included in the DGP. A brief illustration in the form of causal directed acyclic graphs is shown in figure~\ref{fig::dag_scenarios}. The different scenarios were created by adjusting the causal beta coefficients in the probability functions \ref{eq::P_A}, \ref{eq::P_L}, \ref{eq::P_M}, \ref{eq::P_Y_cox} and \ref{eq::P_Y_aalen}, as summarised below. The scenarios are:

\begin{description}
	\item[\textbf{Scenario 1}] $A$ is randomly assigned, meaning there are no confounders of the treatment - mediator or treatment - outcome relationships. Additionally, $M_t$ is only causally influenced by $A$ and $X$, meaning there is also no time-dependent confounding of the mediator - outcome relationship. The only considered time-dependent covariate $L_t$ is a pure predictor ot $Y_{t+1}$, meaning it is not causally influenced by $A$ or $M_{t}$. The only real confounder of the mediator - outcome relationship is $X$. This is the simplest baseline scenario, which all other scenarios are built upon. The specific values used were: $\beta_{0}^{A} = \log(0.5)$ and $\beta_{Z}^{A} = \beta_{A}^{L} = \beta_{M}^{L} = \beta_{Z}^{M} = \beta_{L}^{M} = \beta_{Z}^{Y} = 0$.
	\item[\textbf{Scenario 2}] The same as scenario 1, with the only difference being that $L_t$ now also has a direct causal effect on $M_{t+1}$ and is thus a time-dependent confounder of the mediator - outcome relationship. The only changed value is $\beta_{L}^{M}$, with $\beta_{L}^{M} = \log(3)$.
	\item[\textbf{Scenario 3}] The same as scenario 2, with the only difference being that $L_t$ is now also caused directly by previous values of $M_{t-1}$, resulting in a feedback loop between $L_t$ and $M_t$. Due to this feedback loop, $L_t$ now lies on the causal pathway between $A$ and $Y_t$ through $A \rightarrow M_t \rightarrow L_{t+1} \rightarrow M_{t+2} \rightarrow Y_{t+3}$. The only parameter changed with respect to scenario 2 is $\beta_{M}^{L} = \log(0.6)$.
	\item[\textbf{Scenario 4}] The same as scenario 3, with the only difference being that $L_t$ is now also directly caused by $A$, making $L_t$ a direct mediator as well, which is in a complex feedback loop with the mediator of interest $M_t$. With respect to scenario 3, only $\beta_{A}^{L} = \log(0.6)$ was changed.
	\item[\textbf{Scenario 5}] The same as scenario 1, except that $A$ is no longer randomized, but caused by $Z$, which also has a direct causal influence on $M_t$, making $Z$ a confounder of the treatment - mediator relationship. With respect to scenario 1, the only parameter changes are: $\beta_{0}^{A} = \log(0.445)$, $\beta_{Z}^{A} = \log(0.7)$, and $\beta_{Z}^{M} = \log(0.7)$. 
	\item[\textbf{Scenario 6}] The same as scenario 1, except that $A$ is no longer randomized, but caused by $Z$, which also has a direct causal influence on $Y_t$, making $Z$ a confounder of the treatment - outcome relationship. With respect to scenario 1, the only parameter changes are: $\beta_{0}^{A} = \log(0.445)$, $\beta_{Z}^{A} = \log(0.7)$, and $\beta_{Z}^{Y} = \log(0.7)$.
\end{description}

All scenarios were additionally stratified into a scenario where $M_t$ does mediate the relationship between $A$ and $Y_t$, and in one where it does not. This was achieved by changing the value of $\beta_{A}^{M}$. To make $M_t$ a mediator, $\beta_{A}^{M} = \log(0.2)$ was used, meaning that the probability $P(M_t=1)$ is highly reduced if $A = 1$. In conjunction with the probability of $Y_t$ always being increased by a factor of 5 if $M_{t-1}=1$, a strong indirect effect through $M_t$ is created. In the no-mediation scenario, $\beta_{A}^{M} = 0$ was used instead, keeping the relationship between $M_{t-1}$ and $Y_t$ unchanged. Additionally, the baseline probability of $Y_t$ was also varied over all considered simulation scenarios, using the values $\beta_{0}^{Y} = \log(0.0001)$ (low probability) and $\beta_{0}^{Y} = \log(0.001)$ (high probability) for the Cox / AFT model specification. For the Aalen model specification, $\beta_{0}^{Y} = 0.00014$ (low probability) and $\beta_{0}^{Y} = 0.0014$ (high probability) were used instead.
\par
A large sample size of $n = 2000$ was used for all scenarios, with the simulation being run until $Y_t$ has turned 1 for each simulated individual, approximating a continuous time process, initially without censoring. If $C < T$, the individual was then censored. Here, $\lambda_C$ differs for every simulation parameter combination. Its' values were chosen so that the resulting percentage of right-censored data is approximately 30\% overall in every scenario (see appendix~\ref{appendix::dgp}).

\subsection{Methods}

The Cox, Aalen and AFT models based difference method was then applied to the respective data twice; once including all relevant confounders and once without including either $L_t$ (in scenarios 1 to 4) or $Z$ (in scenarios 5 and 6). Because the actual DGP satisfies the Markov property, only the immediately previous values of $L_t$ and $M_t$ were included as time-varying covariates in the direct effects models, without further inclusion of their histories. All models were fit using time-constant coefficients for all included variables. The AFT models were fit using a Weibull baseline hazard function, approximating the true underlying Cox model \parencite{Carroll2003}.
\par
When analyzing the data in which $Y_t$ follows a Cox / AFT model, we used the following steps to apply the parametric g-formula. First, we fitted pooled log-binomial models for $L_t$, $M_t$ and $Y_t$, if possible. If a model did not converge, a Poisson model with a log-link was fit instead, which generally produces equivalent coefficients \parencite{Chen2018}. These models were then used to simulate a single dataset for the $T^{1,0}$ scenario, which was then pooled with the observed data under $A = 0$ or $A = 1$ as needed and analysed using univariable versions of the Cox or AFT models as outlined in section~\ref{sec::methods_g_form}. In scenario 5 and 6, $Z$ was additionally included in the total effects model. For the data simulation step we used the same discrete-event simulation algorithm as for the true data generation process, only substituting the true coefficients for the estimated ones. Censoring was added by first calculating the maximum likelihood estimate of the rate parameter for the exponential censoring distribution, given the observed censoring times, and then drawing values for $C$ from it. The same strategy was used when analyzing the data in which $Y_t$ follows an Aalen model, with the small exception that an Aalen model was used to model $Y_t$ instead of a log-binomial model.
\par
We calculated total, direct and indirect effects for all methods using $1000$ simulation replications.

\newpage

\subsection{Target estimand calculation} \label{sec::target_calc}

Because we used a discrete-event simulation approach that does not directly specify the marginal causal effects of interest, the true value of the target estimands was not known in all considered simulation scenarios. To calculate the unknown values, we used the method described by \textcite{Naimi2025}, in which the structural equations are adjusted to allow direct simulation of data under different counterfactual scenarios. The simulated data for the relevant scenarios are then put together and analysed using univariable statistical methods, mimicking the approach described in section~\ref{sec::methods_g_form}, with the main difference being that the known true conditional probability functions are used instead of estimated ones.
\par
For the target estimands considered here, we simulated data under three counterfactual scenarios: (1) everyone receiving the treatment $A$, (2) no one receiving the treatment $A$ and (3) everyone receiving treatment $A$, with the mediator process $M_t$ evolving as if no one had received treatment $A$. We generated $n = 1,000,000$ independent observations for each of these three scenarios, using the same random number generation seed for each one. The simulation were run until every individual experienced its first event and were then censored using the same censoring distribution as in the observed data. To calculate the total effect, the datasets from (1) and (2) were pooled together and a univariate Cox, Aalen or AFT model, including only an indicator for which scenario the observation belongs to as an independent variable, was then fit to this data. The beta coefficient of the indicator was then be used as the target estimand of the total effect on the respective scale. The direct effect and indirect effects were calculated similarly, with the only difference being that the datasets from (1) and (3) or (2) and (3) were pooled, respectively. This method was used separately for all considered simulation scenarios and parameter combinations described in section~\ref{sec::scenarios}.

\subsection{Performance criteria} \label{sec::performance}

Our main interest lies in the bias of the indirect effect on the different considered scales. The bias when estimating the direct and total effects was also examined for completeness. We define the bias of each component as the expected difference between the true and estimated values. Formally:

\begin{equation}
	\text{Bias}\big(\hat{\Psi}\big) = \mathbb{E}\big(\Psi - \hat{\Psi}\big),
\end{equation}

where $\Psi$ and $\hat{\Psi}$ are substituted by $\Psi_{TE}, \Psi_{DE}$ and $\Psi_{IE}$ and their estimates on their respective scales. Similarly, the mean squared error is defined as:

\begin{equation}
	\text{MSE}\big(\hat{\Psi}\big) = \mathbb{E}\!\left(\big(\Psi - \hat{\Psi}\big)^2\right),
\end{equation}

which is used to judge the efficiency of the considered methods. Monte-Carlo standard error were calculated by applying the regular standard error equation to the performance criteria over all simulation runs. Because there is no known approximate or exact equation for the standard errors for any of the methods included in this simulation study, type-I error, confidence interval coverages or statistical power are not calculated in this study. In practice, the standard error may be estimated using non-parametric bootstrap methods \parencite{Efron1979}.

\subsection{Computational details}

The entire simulation study was performed using the R programming language (version 4.3.3). The \texttt{simDAG} R package \parencite{Denz2025} was used to generate the required datasets, and the \texttt{survival}, \parencite{Therneau2025} \texttt{timereg} \parencite{Scheike2011} and \texttt{flexsurv} \parencite{Jackson2016} R packages were used to fit the required Cox, Aalen and AFT models with time-dependent covariates, respectively. The R code used is available as part of the online appendix. Details about the data generation procedure are also given in appendix~\ref{appendix::dgp}.

\subsection{Results} \label{sec::results}

\begin{figure}[!htb]
	\centering
	\includegraphics[width=1\linewidth]{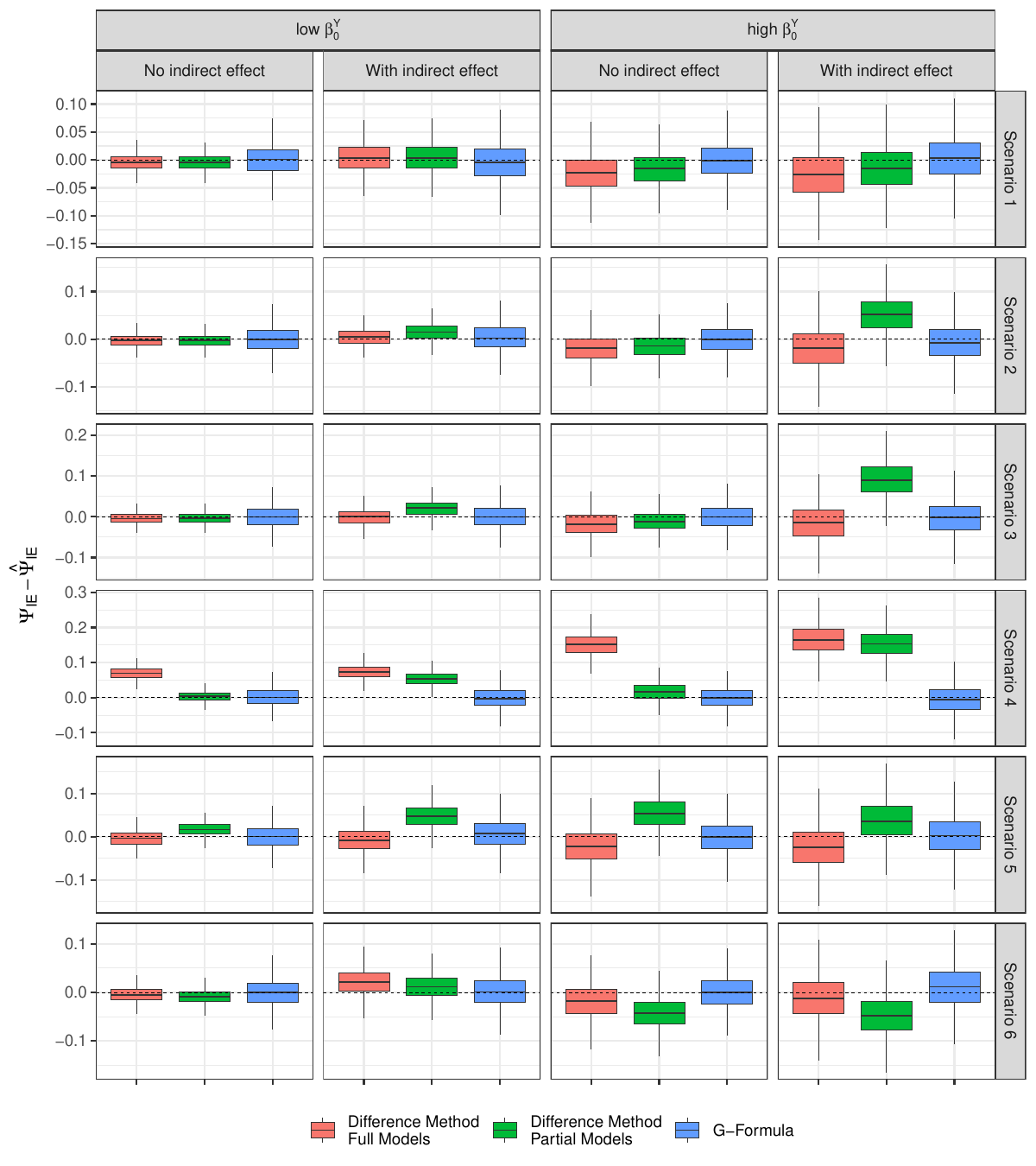}
	\caption{Boxplots displaying the difference between the true indirect effect and the estimated indirect effect on the log hazard ratio scale for the Cox model based difference method and the parametric g-formula for all considered simulation scenarios and parameter combinations (outliers not shown). $\beta_{0}^Y$ is the intercept of the log-binomial model used to generate data for $Y_t$. ``Full models'' refers to Cox models that include all confounders as independent variables, e.g. $X$ and $L_t$ in scenario 1-4 and $X$ and $Z$ in scenarios 5 and 6. ``Partial models'' excluded $L_t$ in scenarios 1-4 and $Z$ in scenarios 5 and 6. All values are based on 1000 simulation repetitions.}
	\label{fig::cox_IE}
\end{figure}

Figure~\ref{fig::cox_IE} shows the distribution of the difference between the true and estimated indirect effect on the log-hazard scale for estimates obtained using both the Cox model based difference method and the g-formula. Here, these methods were applied to the data generated based on a Cox / AFT model. In scenario 1, 2 and 3, when the outcome is rare and $L_t$ was included in the direct effects model, the difference method produced unbiased estimates on average. When $L_t$ is omitted from the model, and the real underlying indirect effect is not 0, this is no longer the case. Conversely, if there is no underlying indirect effect through $M_t$, the exclusion of $L_t$ does not seem to introduce bias. Importantly, if the outcome is not rare, the difference method produces biased estimates regardless of the scenario and the presence of an indirect effect. The estimates obtained using the parametric g-formula, on the other hand, are consistently unbiased in all three scenarios and with all considered parameter combinations.
\par
Furthermore, in scenario 4, the difference method consistently produces biased estimates, regardless of whether $L_t$ is included in the direct effects model or not. The bias is substantially larger if it is included, but it is also present if it is not. The only exception occurs when no indirect effect is present, $L_t$ is omitted from the direct effects model, and the outcome is rare. Again, the parametric g-formula does not show any bias. Similar to the first three scenarios, the difference method show negligible bias on average in scenarios 5, whenever the outcome is rare and the models are correctly specified (e.g. both models including the baseline confounder $Z$ as an independent variable). Otherwise, they are biased, with no substantial difference whether direct or indirect effects are present or not. The results are similar for scenario 6, although a small amount of bias can be seen if an indirect effect is present, even if the outcome is rare. The parametric g-formula remains unbiased in both of these scenarios as well. Further results for the total and direct effects on the log-hazard ratio scale are shown in the online supplement.
\par
\begin{figure}[!htb]
	\centering
	\includegraphics[width=1\linewidth]{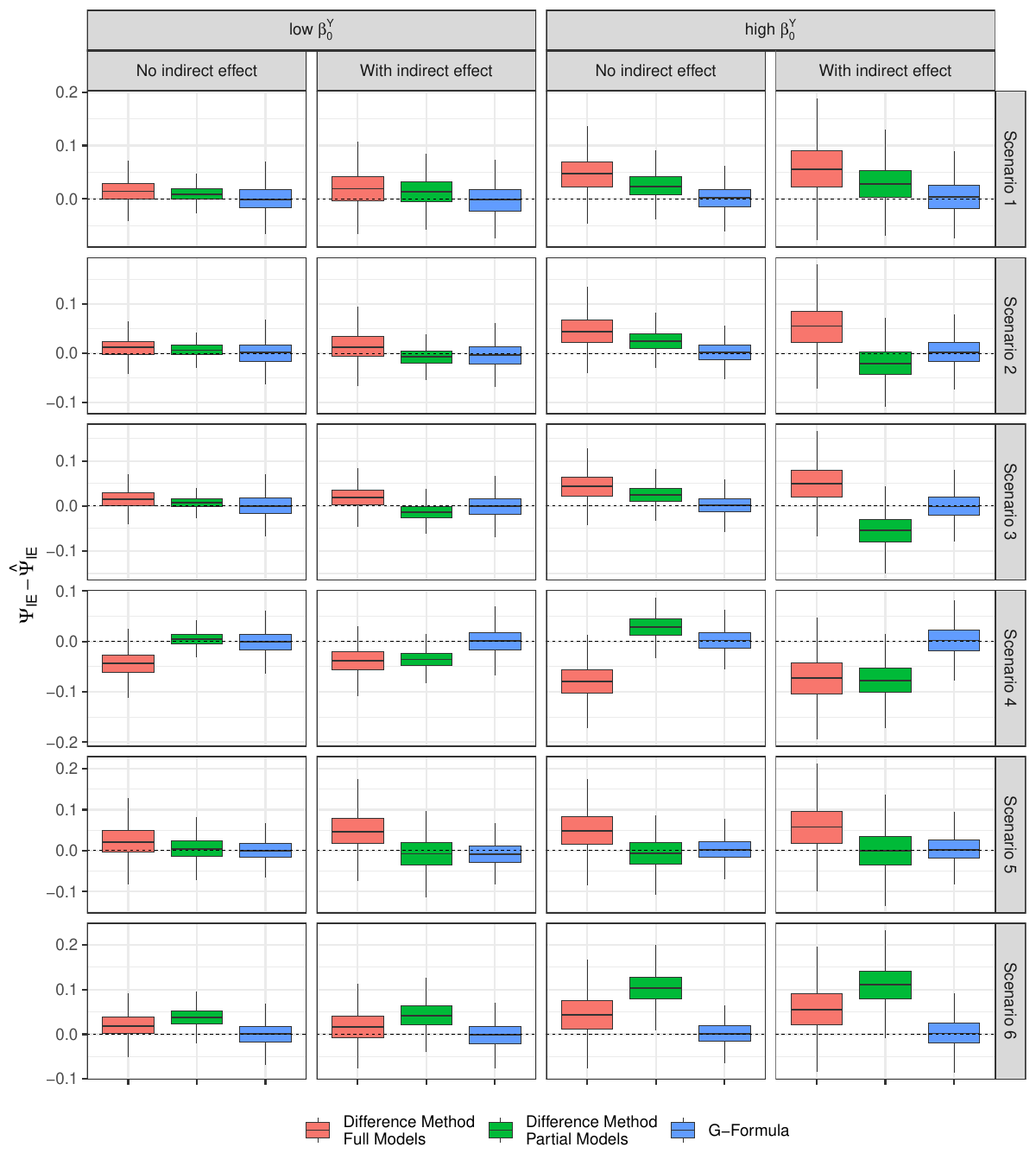}
	\caption{Boxplots displaying the difference between the true indirect effect and the estimated indirect effect on the log mean survial time scale for the AFT model based difference method and the parametric g-formula for all considered simulation scenarios and parameter combinations (outliers not shown). $\beta_{0}^Y$ is the intercept of the log-binomial model used to generate data for $Y_t$. ``Full models'' refers to AFT models that include all confounders as independent variables, e.g. $X$ and $L_t$ in scenario 1-4 and $X$ and $Z$ in scenarios 5 and 6. ``Partial models'' excluded $L_t$ in scenarios 1-4 and $Z$ in scenarios 5 and 6. All values are based on 1000 simulation repetitions.}
	\label{fig::aft_IE}
\end{figure}

Figure~\ref{fig::aft_IE} shows the corresponding results for the indirect effect estimation on the log survival time scale, obtained using the AFT model based difference method and the parametric g-formula. Some similarities to the log hazard ratio scale can be observed. First, as before, the parametric g-formula produces unbiased results on average in all considered scenarios and parameter combinations. Additionally, the bias of the difference method is substantial in scenario 4, especially when $L_t$ is included in the direct effects model. Contrary to the results shown in figure~\ref{fig::cox_IE}, however, the AFT based difference method never produced completely unbiased results. In all scenarios and parameter combinations, a small bias can be observed, with larger bias being observed when baseline probability of the outcome is high. Although across scenarios 1--3, as well as 5 and 6, the bias is generally very small when $L_t$ is included and the outcome is rare, it is never zero.
\par
To investigate the source of this bias further, table~\ref{tab::bias_aft} shows the mean estimates of the total and direct effects obtained from the (full) AFT models in scenario 1 in comparison to the true target estimand value and the true $\beta_A^Y$ used in the DGP (see equation~\ref{eq::P_Y_cox}). Regardless of whether there is an indirect effect through $M_t$ or not, the total effect is not equal to the underlying $\beta_A^Y$. The total effects models are still able to correctly estimate the total effect, as expected. This is no longer the case for the direct effects model. Here, most of the models consistently estimate the true underlying $\beta_A^Y$ instead, leading to a biased estimate of the true marginal direct effect and hence to a biased estimate of the indirect effect. In other words, the marginal effect of interest is not equal to the conditional effect captured by the beta coefficient. As table~\ref{tab::bias_aft} further shows, these results remain consistent in the absence of any right-censoring.
\par
\begin{figure}[!htb]
	\centering
	\includegraphics[width=1\linewidth]{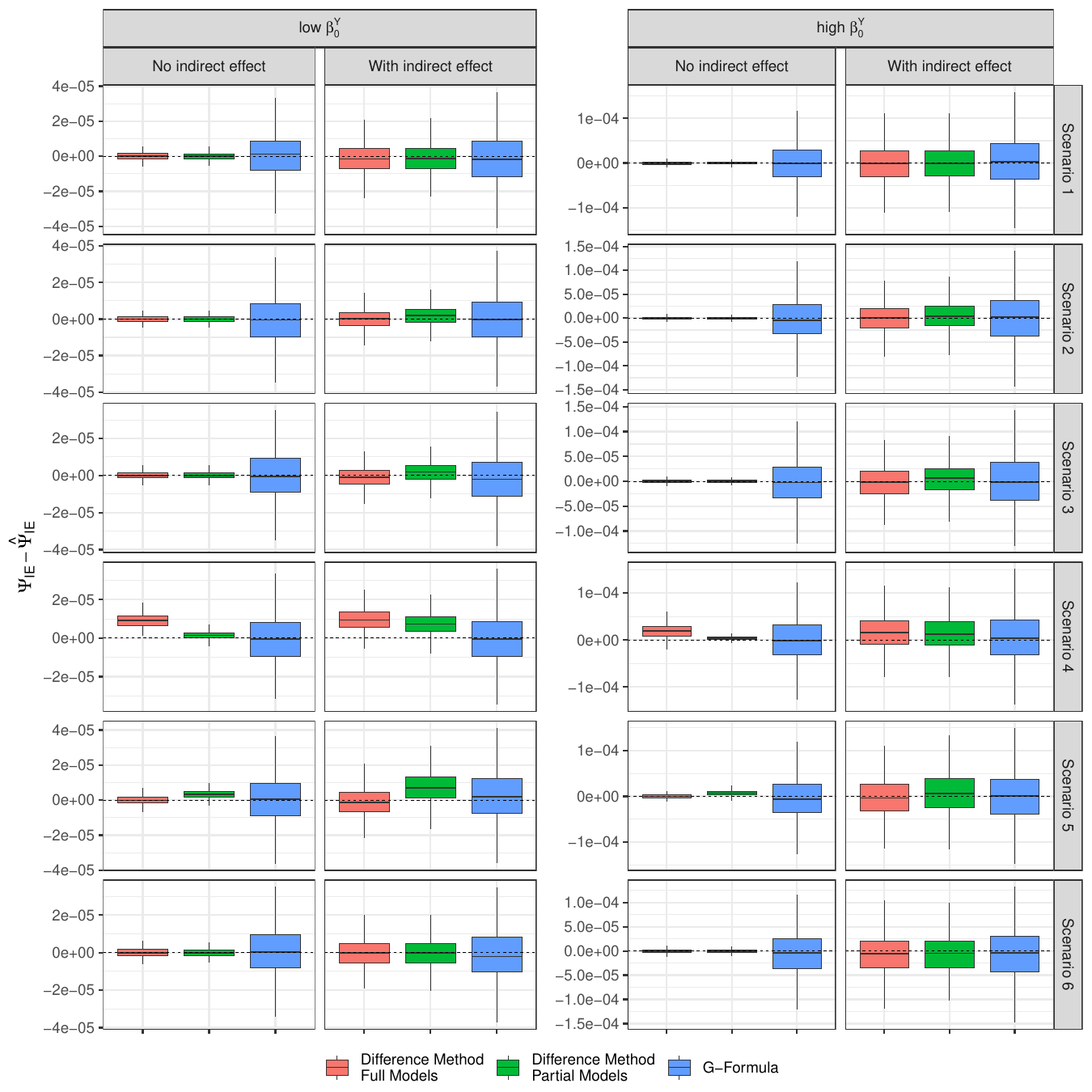}
	\caption{Boxplots displaying the difference between the true indirect effect and the estimated indirect effect on the hazard difference scale for the Aalen model based difference method and the parametric g-formula for all considered simulation scenarios and parameter combinations (outliers not shown). $\beta_{0}^Y$ is the intercept used when generating data for $Y_t$ using the Aalen model DGP. ``Full models'' refers to Aalen models that include all confounders as independent variables, e.g. $X$ and $L_t$ in scenario 1-4 and $X$ and $Z$ in scenarios 5 and 6. ``Partial models'' excluded $L_t$ in scenarios 1-4 and $Z$ in scenarios 5 and 6. All values are based on 1000 simulation repetitions.}
	\label{fig::aalen_IE}
\end{figure}

Finally, figure~\ref{fig::aalen_IE} displays the distribution of the difference between the true and estimated indirect effect on the hazard difference scale, as estimated using the Aalen model based difference method and the g-formula. In contrast to the results shown above, the results in this figure were obtained by applying these methods to the data in which the outcome was generated according to a true Aalen model. The results in figure~\ref{fig::aalen_IE} are very similar to the ones obtained for the Cox model based difference method with a low baseline event probability. The main difference is that the Aalen model based difference method also produces unbiased estimates when the baseline event probability is high. Again, the g-formula also produces unbiased estimates on average in all scenarios and parameter combinations.
\par
Across all scenarios, parameter combinations and scales, whenever the difference method produced unbiased estimates, the mean-squared error of these estimates was lower than the mean-squared error of the corresponding parametric g-formula. This can be seen directly in figures~\ref{fig::cox_IE}, \ref{fig::aft_IE} and \ref{fig::aalen_IE}, by the larger spread of the bias. This difference in performance is most evident on the hazard difference scale with a true underlying Aalen model. A contributing factor to this larger variance is that our implementation of the parametric g-formula uses simulations to obtain the point estimate. To keep this simulation study feasible, we only simulated one dataset corresponding to the size of the input dataset to obtain the point estimates with this method. In real applications, multiple replications would be used and the point estimate would be calculated as the average of these replications, likely reducing the variance substantially.
\par
Additional results, in which the Cox and AFT model based difference method was applied to the data generated according to an Aalen additive hazards model and vice versa to study the impact of model misspecification, are given in the online supplement. Furthermore, tables including the average bias, mean-squared error and their associated Monte-Carlo standard errors for all methods, scenarios and estimands are also given in that online supplement. Finally, to demonstrate the effect of right-censoring on the results, the online supplement additionally includes a repetition of the entire analysis without any right-censoring.

\begin{landscape}
	\begin{table}[!htb]
		\centering
		\begin{tabular}{llcccccc}
			\toprule
			Estimand & Kind & $\log\!\left(\beta_{0}^Y\right)$ & \% right-censored & Mean estimate & MCSE & True $\beta_{A}^Y$ & True target estimand \\
			\midrule
			\multirow{8}{*}{Total effect} & \multirow{4}{*}{No indirect effect} & \multirow{2}{*}{0.0001} & 0 & 0.094 & 0.001 & 0.105 & 0.094 \\
			& & & 30 & 0.086 & 0.001 & 0.105 & 0.089 \\
			\cmidrule{3-8}
			& & \multirow{2}{*}{0.001} & 0 & 0.058 & 0.001 & 0.105 & 0.059 \\
			& & & 30 & 0.059 & 0.001 & 0.105 & 0.058 \\
			\cmidrule{2-8}
			& \multirow{4}{*}{With indirect effect} & \multirow{2}{*}{0.0001} & 0 & 0.319 & 0.002 & 0.105 & 0.317 \\
			& & & 30 & 0.346 & 0.001 & 0.105 & 0.346 \\
			\cmidrule{3-8}
			& & \multirow{2}{*}{0.001} & 0 & 0.609 & 0.001 & 0.105 & 0.608 \\
			& & & 30 & 0.565 & 0.001 & 0.105 & 0.567 \\
			\midrule
			\multirow{8}{*}{Direct effect} & \multirow{4}{*}{No indirect effect} & \multirow{2}{*}{0.0001} & 0 & 0.106 & 0.002 & 0.105 & 0.094 \\
			& & & 30 & 0.099 & 0.002 & 0.105 & 0.085 \\
			\cmidrule{3-8}
			& & \multirow{2}{*}{0.001} & 0 & 0.104 & 0.002 & 0.105 & 0.059 \\
			& & & 30 & 0.106 & 0.002 & 0.105 & 0.060 \\
			\cmidrule{2-8}
			& \multirow{4}{*}{With indirect effect} & \multirow{2}{*}{0.0001} & 0 & 0.109 & 0.003 & 0.105 & 0.093 \\
			& & & 30 & 0.107 & 0.002 & 0.105 & 0.089 \\
			\cmidrule{3-8}
			& & \multirow{2}{*}{0.001} & 0 & 0.116 & 0.002 & 0.105 & 0.059 \\
			& & & 30 & 0.115 & 0.002 & 0.105 & 0.058 \\
			\bottomrule
		\end{tabular}
		\medskip
		\caption{Mean estimate and associated Monte-Carlo standard errors (MCSE) for the estimation of the beta coefficients in the full direct effects accelerated failure time (AFT) model and the AFT total effects model in simulation scenario 1. The true beta coefficients for the AFT model are equal to the negative of the known beta coefficients of the underlying data generation process (DGP). All values are rounded to the third decimal point. All values are based on 1000 simulation repetitions.}
		\label{tab::bias_aft}
	\end{table}
\end{landscape}

\FloatBarrier
\newpage 

\section{Results for the motivating example} \label{sec::example}

Below we describe the analysis of the motivating example introduced in section~\ref{sec::example_introduction}. We first applied the difference method based on Cox, Aalen and AFT models to the data. The outcome ($Y_t$) was the first date until either a liver transplant or death occurred, which happened for 28 patients in total. The suspected mediator ($M_t$), histological progression by two stages, was included as a time-dependent covariate which was set to 0 before the date at which the progression occurred and to 1 afterwards. Because the treatment with UDCA ($A$) was completely randomized in this study, there is by definition no confounding of the relationship between UDCA and LTFS or UDCA and histological progression. However, there may be confounding of the causal relationship between histological progression and LTFS.
\par
As exemplary baseline adjustment variables, we included the stage of the disease at randomization, the Mayo PBC risk score \parencite{Jacob2008} at randomization and the natural logarithm of the bilirubin value at randomization, which is an important biomarker for liver function \parencite{Trivella2023}, in the direct effects models. As potential baseline confounders, these variables play the same role as $X$ in the rest of the article. Multiple potential time-varying confounders were measured in the study. These include the appearance of esphogeal varices, ascites and encephalopathy, as well as the doubling of of initial bilirubin levels. All of these variables could be argued to play the role of $L_t$ in the analysis. Due to the small sample size and co-linearity among some of these variables, it is not possible to include all of them in the analysis. Instead, only the appearance of varices was included as a time-varying covariate, using the same coding scheme as for the histological progression. The Aalen models were fit using time-varying coefficients for all variables, because the proportional hazards assumption was found to be appropriate for this dataset, \parencite{Lindor1994, Therneau2000} which implies a time-varying hazard difference. For the AFT model, we used a Weibull baseline distribution.
\par
For the parametric g-formula, the dataset was first transformed into the long-format, including a single row per day under observation for each individual. Three logistic regression models were then fit to the data: one for LTFS, one for histological progression and one for appearance of varices. Each model only contained data up to the first time that the variable turned 1. All models included the three baseline variables, with the model for LTFS also including the previous values of the appearance of varices and histological progression. The model for the appearance of varices included previous values of histological progression and vice versa. These models were then used to apply the parametric g-formula as described in section~\ref{sec::methods_g_form}. For the main analysis, the g-formula was applied 10 times and the average of the results was used as a point estimate. All analyses were repeated using patient-level non-parametric bootstrapping with 1000 repetitions. 95\% confidence intervals were derived from these bootstrap samples using the percentile method \parencite{Efron1979}. The R code used for the analysis is available as part of the online appendix.
\par

\begin{table}[!htb]
	\centering
	{\footnotesize
	\begin{tabular}{llcccc}
		\toprule
		& \multirow{2}{*}{\textbf{Estimand}} & \multicolumn{2}{c}{\textbf{Hazard Ratio Scale}} & \multicolumn{2}{c}{\textbf{Survival Time Ratio Scale}} \\[3pt]
		
		& & Estimate & 95\% CI & Estimate & 95\% CI \\
		\midrule
		& Total Effect & 0.63 & $[0.28, 1.39]$ & 1.41 & $[0.80, 2.56]$ \\[3pt]
		Difference Method & Direct Effect & 0.50 & $[0.17, 1.14]$ & 1.43 & $[0.95, 2.99]$ \\[3pt]
		& Indirect Effect & 1.25 & $[0.81, 2.64]$ & 0.99 & $[0.57, 1.28]$ \\[3pt]
		\midrule
		& Total Effect & 0.62 & $[0.26, 1.42]$ & 1.41 & $[0.80, 2.84]$ \\[3pt]
		Parametric G-Formula & Direct Effect & 0.55 & $[0.23, 1.21]$ & 1.81 & $[0.86, 4.13]$ \\[3pt]
		& Indirect Effect & 1.14 & $[0.68, 1.97]$ & 0.86 & $[0.52, 1.34]$ \\[3pt]
		\bottomrule
	\end{tabular}}
	\medskip
	\caption{Estimates of the total, direct and indirect effects of UDCA on LTFS with respect to histological progression by two stages on the hazard ratio and survival time ratio scale. The difference method based estimates for the hazard ratio scale were obtained using two Cox models, and the corresponding estimates for the survival time ratio scale were obtained using two accelerated failure time models. The associated 95\% confidence intervals were obtained using 1000 patient-level bootstrap samples and the percentile method. All values were rounded to the second digit.}
	\label{tab::example_results_tc}
\end{table}

\begin{figure}[!htb]
	\centering
	\includegraphics[width=0.8\linewidth]{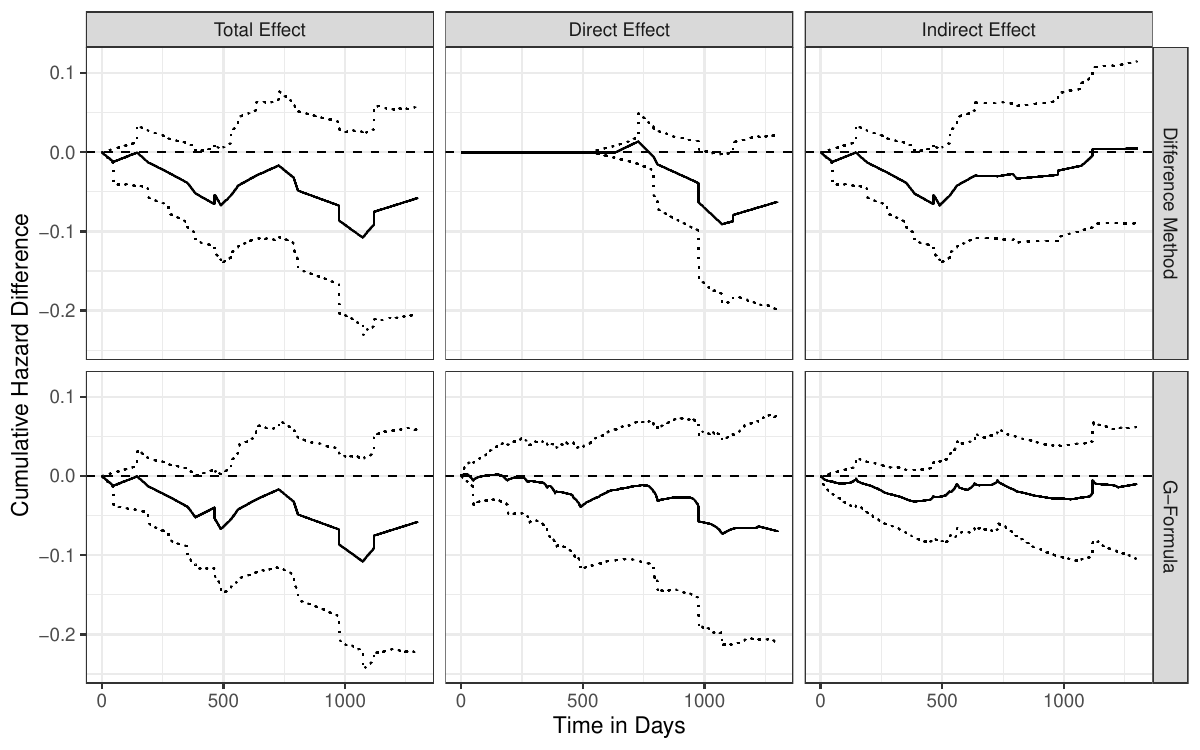}
	\caption{Estimates of the time-dependent total, direct and indirect effects of UDCA on LTFS with respect to histological progression by two stages on the cumulative hazard difference scale, as obtained using the Aalen model based difference method and the parametric g-formula (solid lines). The associated 95\% confidence intervals (dotted lines) were obtained using 1000 patient-level bootstrap samples and the percentile method.}
	\label{fig::example_haz_diff}
\end{figure}

Table~\ref{tab::example_results_tc} shows the point estimates and 95\% confidence intervals of the total, direct and indirect effects of UDCA with respect to histological progression that were obtained using the difference method and the g-formula on both the hazard ratio and survival time ratio scale. Across both scales and both analysis methods, the results are fairly consistent. Estimates of the total effect of UDCA on LTFS consistently show a protective effect of UDCA, with a hazard ratio of approximately 0.63 or a survival time ratio of 1.41. Due to the small sample size, however, all associated 95\% confidence intervals include 1. The direct effect estimates are very similar to the total effect estimates, while the indirect effect estimates are very close to 1. Again, all 95\% confidence intervals contain the reference value 1. Figure~\ref{fig::example_haz_diff} shows similar results as table~\ref{tab::example_results_tc}, but on the time-dependent cumulative hazard difference scale, as obtained using the Aalen model based difference method and the parametric g-formula. The results are similar to the ones obtained on the other scales. Again, no substantial difference between the methods can be observed.
\par
This analysis has multiple limitations. First, the small sample size clearly inhibits our ability to draw substantial conclusions, highlighting the importance of performing adequate sample size calculations if the goal is to perform mediation analysis. In most cases, a randomized trial that has adequate statistical power to detect the main effect will not have adequate statistical power for such analyses \parencite{VanderWeele2015}. Secondly, multiple required assumptions for both methods may be violated here. Although the event is rare and the data seem consistent with a Cox proportional hazards model \parencite{Lindor1994, Therneau2000}, it is doubtful that UDCA has no direct effect on any of the potential time-dependent confounders, such as the development of varices. If such effects exist, the difference method should not be used, as described throughout the article. Developing a causal DAG based on expert opinion and the available literature would be required to assess this assumption more formally.
\par
Regardless of the presence of such structural issues, it is also highly doubtful that the sequential ignorability assumption (see section~\ref{sec::assumptions}) holds here in general. Due to the sample size limitations, we were not able to adjust for all available time-varying variables, that might act as confounders of the mediator - outcome relationship in either analysis. Since this assumption is also needed for the g-formula, the results of both methods are potentially affected. Further, we were unable to investigate whether an interaction between UDCA and histological progression is present, or whether censoring truly is independent, conditional on the included covariates. The violation of either assumptions could also introduce bias in both methods, which may explain the near equivalence of the estimates produced by the g-formula and the difference method.

\FloatBarrier

\section{Discussion} \label{sec::discussion}

In this article we investigated the performance of the difference method for causal mediation analysis in the presence of a time-dependent mediator, time-dependent confounders and a time-to-event outcome. We performed a comprehensive simulation study, considering different kinds of scenarios that may arise in real data analyses. The results were highly dependent on the underlying DGP and the employed models. In simulation scenarios 1--3, where $L_t$ was merely an independent predictor of $Y_t$ (scenario 1), an otherwise independent time-dependent confounder of the mediator outcome relationship (scenario 2) or even a time-dependent confounder in a feedback loop with the mediator (scenario 3), the difference method was unbiased when Aalen models with time-dependent covariates were used. The addition of baseline confounders of the treatment - mediator (scenario 5) or treatment - outcome relationship (scenario 6) did not change these results. Results for the Cox model were similar when the event was relatively rare. However, if the baseline hazard of the event was increased, biased estimates were produced, mirroring previous results considering only time-fixed mediators \parencite{Lange2011, Fulcher2017}. These results were consistent, regardless of whether an actual indirect effect through $M_t$ was present or not.
\par
The AFT model based difference method showed slightly different results. Very small but consistent bias was found in nearly all scenarios and parameter combinations. This unexpected result was not due to simulation error, convergence issues or classic model misspecification, since the DGP used for this analysis can be described as a Cox model with a constant baseline hazard, which is mathematically consistent with an AFT model based on a Weibull distribution \parencite{Carroll2003}. It was also not the result of model misspecification due to censoring \parencite{Fulcher2017, Ochoa2020}, since an analysis without any censoring showed similar results. Instead, the reason was non-collapsibility of the beta coefficients. The direct effects model consistently estimated the conditional beta coefficient of the true DGP, but the true marginal direct effect was not equal to this coefficient, even when no confounding was present. This result is in contrast to previous studies dealing with time-fixed mediators, in which the coefficients of AFT models were found to be collapsible \parencite{Crowther2023}.
\par
Importantly, all versions of the difference method produced highly biased estimates when the treatment had a direct causal effect on the time-dependent confounder. The reason for this bias is, that the direct effect of $A$ on $Y_t$ with respect to $M_t$ cannot be identified by standard models, because $L_t$ is both a confounder for the mediator - outcome relationship and a mediator for the treatment - outcome relationship. Conditioning on $L_t$ correctly adjusts for the former, but also incorrectly adjusts for the latter. Conversely, not conditioning on $L_t$ correctly leaves open the mediating pathways, but fails to adjust for the confounding ones. This problem was less pronounced in scenarios without indirect effects of $M_t$, but the bias was not zero. This suggests that, contrary to situations with time-fixed mediators and a time-to-event outcome, the difference method cannot generally be used to test for the presence of time-dependent mediation, without making further assumptions \parencite{VanderWeele2011}. The parametric g-formula on the other hand produced unbiased estimates in all scenarios, regardless of the underlying DGP.
\par
Our results suggest that, under the stated identifiability and model assumptions, the Cox or Aalen based difference method, but not the AFT model based difference method, may be used when there are no time-dependent confounders that are directly caused by the treatment. It is, however, important to note that the presented simulations do not constitute a mathematical proof of this result. Although we used a large amount of different settings and scenarios, it is possible that the method behaves differently under extreme parameter values or slightly different DGPs. For example, although our simulations showed negligible non-collapsibility issues for the Cox model based difference method in most scenarios with a rare outcome, exactly how rare the outcome needs to be remains unclear. Additionally, although many scenarios were covered, we did not consider all complications that may arise in real-world analysis of time-to-event data. For example, we did not consider interactions between the mediator and the treatment variable \parencite{Yan2025}, the combined indirect effect of multiple mediators \parencite{Lange2014, Huang2025}, dependent censoring \parencite{Fulcher2017}, competing risks \parencite{DomingoRelloso2026}, various forms of truncation \parencite{Yu2025} or measurement error \parencite{Bhandari2025, VanderWeele2013}, all of which would further complicate the analysis.
\par
It is also important to consider censoring carefully. In this article, we defined the target estimand without additional interventions on censoring events, meaning that the target estimand is implicitly dependent on the censoring mechanism as well \parencite{Wen2025}. In some cases, researchers may instead be interested in the direct or indirect effects that would be observed in the absence of censoring. This may complicate the use of the classical difference method further. First, if censoring is not independent, adjustment for dependent censoring is necessary, but not always possible \parencite{Fulcher2017}. Secondly, if an indirect effect through $A \rightarrow M_t \rightarrow T$ exists, the total effect of $A$ on $T$ is time-dependent whenever the distribution of $M_t$ changes over time. For example, in one of the DGP used in our simulation study, a time-constant hazard ratio for $A$ on $T$ was used, but since $M_t$ slowly turns 1 for every individual, partially through the influence of $A$, the total effect varies over time. The beta coefficient in the total effects Cox model, for instance, is then only an \emph{average} over the observed follow-up time. Right-censoring shortens this time, which would result in the coefficient targeting a different average than the un-censored true average. In such cases, the use of time-dependent coefficients for $A$ in the total effects model would thus be necessary.
\par
We also focused solely on a binary baseline treatment variable. If the treatment changes over time, even more complex forms of feedback between the time-dependent variables are possible. Similar to the discussed issue of estimating the direct effect of $A$ on $Y_t$ in the presence of a direct causal relationship between $A$ and $L_t$, the causal effect of $A$ on $M_t$ or of $A$ on $Y_t$ is usually not identified by simple conditioning, if $L_t$ or $M_t$ act as both confounders and mediators of the respective relationship \parencite{Hernan2020}. Therefore, the difference method would likely fail in these scenarios. In comparison, under the assumptions used throughout the study, the parametric g-formula would still be able to produce unbiased estimates, as shown in the literature \parencite{Lin2017a}.
\par
In general, the difference method should only be used very carefully with a time-dependent mediator and a time-to-event outcome. All required assumptions, including the regular causal identifiability assumptions, model based assumptions (such as the proportional hazards assumption) and structural assumptions should be assessed carefully with respect to the respective study question. These assumptions can be considered restrictive or even ``heroic'' \parencite{Hernan2020} in practice, but may hold in some cases. For example, the required assumption that none of the time-dependent confounders are themselves directly caused by the treatment of interest may be hard to defend when considering the effect of UDCA on survival, but might be plausible when trying to estimate the indirect effect of a Herpes Zoster vaccination on the development of dementia, through the prevention of Herpes Zoster manifestations \parencite{Lophatananon2023, Eyting2025}. If those assumptions can be reasonably defended, the difference method may be used as a simple and efficient alternative to more complex methods. This may be especially helpful in situations with many discrete points in time or continuous time, because most other methods are not designed for such scenarios, with some notable exceptions \parencite{Kateline2025, Bhandari2025}. In most cases, however, the usage of specifically designed methods that require less stringent structural assumptions, such as the parametric mediational g-formula, \parencite{Lin2017a} is preferable.
\par
Although an increasing amount of such methods are being proposed in the literature \parencite{Kormaksson2024, Lin2017, Lin2017a, Vansteelandt2019, Kateline2025, Aalen2020, Breum2024, Huang2021, Gao2023, Valeri2023, Bhandari2025, Bhandari2025a, Zeng2023, Wang2025}, it is currently unclear which is the most appropriate in which scenario. Future research should focus on systematically reviewing and comparing these methods and their associated target estimands before developing further methods. Both a neutral comparison study \parencite{Boulesteix2017} and user-friendly software implementations are needed to facilitate their usage by applied researchers. We hope that the simulation study design presented in this article may be useful for such future studies.

\section*{Acknowledgments}

We would like to thank all members of the Department of Medical Informatics, Biometry and Epidemiology at the Ruhr-University Bochum for their valuable input and the multiple discussions about the contents of this paper.

\section*{Financial disclosure}

None reported.

\section*{Conflict of interest}

The authors declare no potential conflict of interests.

\section*{Data availability statement}

The data for the study on Ursodeoxycholic acid in primary biliary cholangitis patients by \textcite{Lindor1994} is freely available as part of the \texttt{survival} R package (\url{https://cran.r-project.org/package=survival}). The R code used for the simulation study and the illustrative example is available as part of the online appendix.

\FloatBarrier
\newpage

\printbibliography

\appendix

\newpage

\section{Target estimand on different scales}

In the main text, the definition for the target estimands were given only on the log-hazard ratio scale. On the hazard difference scale the estimands are defined as:

\begin{align}
	\Psi_{TE}(t) = & \lambda^{1, 1}(t) - \lambda^{0, 0}(t), \label{eq::target_hz_diff_TE} \\
	\Psi_{DE}(t) = & \lambda^{1, 0}(t) - \lambda^{0, 0}(t), \label{eq::target_hz_diff_DE} \\
	\Psi_{IE}(t) = & \lambda^{1, 0}(t) - \lambda^{1, 1}(t). \label{eq::target_hz_diff_IE}
\end{align}

Similarly, on the log survival time ratio scale, the estimands are defined as:

\begin{align}
	\Psi_{TE} = & \mathbb{E}\left(\log\!\left(T^{1, 1}\right) - \log\!\left(T^{0, 0}\right)\right), \label{eq::target_logT_TE} \\
	\Psi_{DE} = & \mathbb{E}\left(\log\!\left(T^{1, 0}\right) - \log\!\left(T^{0, 0}\right)\right), \label{eq::target_logT_DE} \\
	\Psi_{IE} = & \mathbb{E}\left(\log\!\left(T^{1, 0}\right) - \log\!\left(T^{1, 1}\right)\right). \label{eq::target_logT_IE}
\end{align}

\section{Data generation process} \label{appendix::dgp}

To generate data corresponding to the DGP described in section~\ref{sec::dgp}, we used the following discrete-event simulation based approach:

\begin{enumerate}
	\item Generate $n$ random draws from a standard normal distribution to generate $X$.
	\item Generate $n$ random draws from a Bernoulli distribution to generate $Z$.
	\item Generate $n$ random draws from a log-binomial regression model based on $Z$ to generate $A$.
	\item Set $t = 0$ for all $n$ individuals.
	\item Initialize three vectors of size $n$ that are filled with zeros for $L_0$, $M_0$ and $Y_0$, representing their values at $t = 0$.
	\item Calculate the probabilities for $L_t$, $M_t$ and $Y_t$ turning one if $t$ is increased by 1, based on equations~\ref{eq::P_L}, \ref{eq::P_M} and \ref{eq::P_Y_cox} or \ref{eq::P_Y_aalen}, using the current state of these variables and the values generated in steps 1-3.
	\item For each of the $n$ individuals and for each of $L_t$, $M_t$ and $Y_t$ that are still not 1, draw a random value from a corresponding left-truncated exponential distribution. The rate of this distribution is set to the corresponding individual-specific probability calculated in step 6 and the left-truncation is the current individual-specific simulation time.
	\item Round up all generated times to the next integer.
	\item For each of the $n$ individuals, choose the smallest of the times generated in the previous step, denoted $t_{next}$.
	\item For each of the $n$ individuals, set the variable(s) corresponding to $t_{next}$ to 1.
	\item For each of the $n$ individuals, set the individual-specific $t$ to $t_{next}$.
	\item For each of the $n$ individuals, repeat steps 6-11 until $Y_t$ turns 1.
\end{enumerate}

\begin{landscape}
	
	\begin{table}[!htb]
		\centering
		{\footnotesize
		\begin{tabular}{ccccccccccc}
			\toprule
			\textbf{Scenario} & $\beta_A^M$ & $\beta_0^Y$ & $\beta^A_0$ & $\beta_Z^A$ & $\beta_A^L$ & $\beta_M^L$ & $\beta_Z^M$ & $\beta_L^M$ & $\beta_Z^Y$ & $\lambda_C$ \\
			\midrule
			
			\multirow{4}{*}{\textbf{1}} & \multirow{2}{*}{0} & $\log(0.0001)$ & $\log(0.5)$ & 0 & 0 & 0 & 0 & 0 & 0 & 0.0004253625 \\
			& & $\log(0.001)$ & $\log(0.5)$ & 0 & 0 & 0 & 0 & 0 & 0 & 0.0021502414 \\
			\cmidrule{2-11}
			& \multirow{2}{*}{$\log(0.2)$} & $\log(0.0001)$ & $\log(0.5)$ & 0 & 0 & 0 & 0 & 0 & 0 & 0.0003650604 \\
			& & $\log(0.001)$ & $\log(0.5)$ & 0 & 0 & 0 & 0 & 0 & 0 & 0.0016433996 \\
			
			\midrule
			
			\multirow{4}{*}{\textbf{2}} & \multirow{2}{*}{0} & $\log(0.0001)$ & $\log(0.5)$ & 0 & 0 & 0 & 0 & $\log(3)$ & 0 & 0.0004374994 \\
			& & $\log(0.001)$ & $\log(0.5)$ & 0 & 0 & 0 & 0 & $\log(3)$ & 0 & 0.0024057371 \\
			\cmidrule{2-11}
			& \multirow{2}{*}{$\log(0.2)$} & $\log(0.0001)$ & $\log(0.5)$ & 0 & 0 & 0 & 0 & $\log(3)$ & 0 & 0.0004104836 \\
			& & $\log(0.001)$ & $\log(0.5)$ & 0 & 0 & 0 & 0 & $\log(3)$ & 0 & 0.0019531746 \\
			
			\midrule
			
			\multirow{4}{*}{\textbf{3}} & \multirow{2}{*}{0} & $\log(0.0001)$ & $\log(0.5)$ & 0 & 0 & $\log(0.6)$ & 0 & $\log(3)$ & 0 & 0.0004263690 \\
			& & $\log(0.001)$ & $\log(0.5)$ & 0 & 0 & $\log(0.6)$ & 0 & $\log(3)$ & 0 & 0.0023093365 \\
			\cmidrule{2-11}
			& \multirow{2}{*}{$\log(0.2)$} & $\log(0.0001)$ & $\log(0.5)$ & 0 & 0 & $\log(0.6)$ & 0 & $\log(3)$ & 0 & 0.0004036540 \\
			& & $\log(0.001)$ & $\log(0.5)$ & 0 & 0 & $\log(0.6)$ & 0 & $\log(3)$ & 0 & 0.0019126007 \\
			
			\midrule
			
			\multirow{4}{*}{\textbf{4}} & \multirow{2}{*}{0} & $\log(0.0001)$ & $\log(0.5)$ & 0 & $\log(0.6)$ & $\log(0.6)$ & 0 & $\log(3)$ & 0 & 0.0004136372 \\
			& & $\log(0.001)$ & $\log(0.5)$ & 0 & $\log(0.6)$ & $\log(0.6)$ & 0 & $\log(3)$ & 0 & 0.0021736180 \\
			\cmidrule{2-11}
			& \multirow{2}{*}{$\log(0.2)$} & $\log(0.0001)$ & $\log(0.5)$ & 0 & $\log(0.6)$ & $\log(0.6)$ & 0 & $\log(3)$ & 0 & 0.0003916444 \\
			& & $\log(0.001)$ & $\log(0.5)$ & 0 & $\log(0.6)$ & $\log(0.6)$ & 0 & $\log(3)$ & 0 & 0.0017965080 \\
			
			\midrule
			
			\multirow{4}{*}{\textbf{5}} & \multirow{2}{*}{0} & $\log(0.0001)$ & $\log(0.445)$ & $\log(0.7)$ & 0 & 0 & $\log(0.7)$ & 0 & 0 & 0.0004112966 \\
			& & $\log(0.001)$ & $\log(0.445)$ & $\log(0.7)$ & 0 & 0 & $\log(0.7)$ & 0 & 0 & 0.0020159880 \\
			\cmidrule{2-11}
			& \multirow{2}{*}{$\log(0.2)$} & $\log(0.0001)$ & $\log(0.445)$ & $\log(0.7)$ & 0 & 0 & $\log(0.7)$ & 0 & 0 & 0.0003272549 \\
			& & $\log(0.001)$ & $\log(0.445)$ & $\log(0.7)$ & 0 & 0 & $\log(0.7)$ & 0 & 0 & 0.0014279447 \\
			
			\midrule
			
			\multirow{4}{*}{\textbf{6}} & \multirow{2}{*}{0} & $\log(0.0001)$ & $\log(0.445)$ & $\log(0.7)$ & 0 & 0 & 0 & 0 & $\log(0.7)$ & 0.0003564597 \\
			& & $\log(0.001)$ & $\log(0.445)$ & $\log(0.7)$ & 0 & 0 & 0 & 0 & $\log(0.7)$ & 0.0019138197 \\
			\cmidrule{2-11}
			& \multirow{2}{*}{$\log(0.2)$} & $\log(0.0001)$ & $\log(0.445)$ & $\log(0.7)$ & 0 & 0 & 0 & 0 & $\log(0.7)$ & 0.0003018970 \\
			& & $\log(0.001)$ & $\log(0.445)$ & $\log(0.7)$ & 0 & 0 & 0 & 0 & $\log(0.7)$ & 0.0013732391 \\
			
			\bottomrule 
		\end{tabular}}
		\medskip
		\caption{An overview of all combinations of parameters used in each scenario of the simulation study, when simulating data for $Y_t$ based on a Cox or AFT model. Simulation parameters that were fixed over all scenarios are not included. $\lambda_C$ was determined by first simulating a single dataset with $1,000,000$ individuals from the respective DGP and then using the Newton-Raphson algorithm to numerically determine the rate needed for an exponential distribution that leads to approximately 30\% censoring.}
		\label{tab::dgp_params_cox}
	\end{table}
\end{landscape}

\begin{landscape}
	
	\begin{table}[!htb]
		\centering
		{\footnotesize
		\begin{tabular}{cccccccccccc}
			\toprule
			\textbf{Scenario} & $\beta_A^M$ & $\beta_0^Y$ & $\gamma$ & $\beta^A_0$ & $\beta_Z^A$ & $\beta_A^L$ & $\beta_M^L$ & $\beta_Z^M$ & $\beta_L^M$ & $\beta_Z^Y$ & $\lambda_C$ \\
			\midrule
			
			\multirow{4}{*}{\textbf{1}} & \multirow{2}{*}{0} & $0.00014$ & $0.00015$ & $\log(0.5)$ & 0 & 0 & 0 & 0 & 0 & 0 & 0.0002077317 \\
			& & $0.0014$ & $0.00015$ & $\log(0.5)$ & 0 & 0 & 0 & 0 & 0 & 0 & 0.0015544418 \\
			\cmidrule{2-12}
			& \multirow{2}{*}{$\log(0.2)$} & $0.00014$ & $0.00015$ & $\log(0.5)$ & 0 & 0 & 0 & 0 & 0 & 0 & 0.0001976828 \\
			& & $0.0014$ & $0.00015$ & $\log(0.5)$ & 0 & 0 & 0 & 0 & 0 & 0 & 0.0013604477 \\
			
			\midrule
			
			\multirow{4}{*}{\textbf{2}} & \multirow{2}{*}{0} & $0.00014$ & $0.00015$ & $\log(0.5)$ & 0 & 0 & 0 & 0 & $\log(3)$ & 0 & 0.0002094211 \\
			& & $0.0014$ & $0.00015$ & $\log(0.5)$ & 0 & 0 & 0 & 0 & $\log(3)$ & 0 & 0.0016280939 \\
			\cmidrule{2-12}
			& \multirow{2}{*}{$\log(0.2)$} & $0.00014$ & $0.00015$ & $\log(0.5)$ & 0 & 0 & 0 & 0 & $\log(3)$ & 0 & 0.0002051545 \\
			& & $0.0014$ & $0.00015$ & $\log(0.5)$ & 0 & 0 & 0 & 0 & $\log(3)$ & 0 & 0.0014590998 \\
			
			\midrule
			
			\multirow{4}{*}{\textbf{3}} & \multirow{2}{*}{0} & $0.00014$ & $0.00015$ & $\log(0.5)$ & 0 & 0 & $\log(0.6)$ & 0 & $\log(3)$ & 0 & 0.0002090750 \\
			& & $0.0014$ & $0.00015$ & $\log(0.5)$ & 0 & 0 & $\log(0.6)$ & 0 & $\log(3)$ & 0 & 0.0016007601 \\
			\cmidrule{2-12}
			& \multirow{2}{*}{$\log(0.2)$} & $0.00014$ & $0.00015$ & $\log(0.5)$ & 0 & 0 & $\log(0.6)$ & 0 & $\log(3)$ & 0 & 0.0002047112 \\
			& & $0.0014$ & $0.00015$ & $\log(0.5)$ & 0 & 0 & $\log(0.6)$ & 0 & $\log(3)$ & 0 & 0.0014473551 \\
			
			\midrule
			
			\multirow{4}{*}{\textbf{4}} & \multirow{2}{*}{0} & $0.00014$ & $0.00015$ & $\log(0.5)$ & 0 & $\log(0.6)$ & $\log(0.6)$ & 0 & $\log(3)$ & 0 & 0.0002066089 \\
			& & $0.0014$ & $0.00015$ & $\log(0.5)$ & 0 & $\log(0.6)$ & $\log(0.6)$ & 0 & $\log(3)$ & 0 & 0.0015535868 \\
			\cmidrule{2-12}
			& \multirow{2}{*}{$\log(0.2)$} & $0.00014$ & $0.00015$ & $\log(0.5)$ & 0 & $\log(0.6)$ & $\log(0.6)$ & 0 & $\log(3)$ & 0 & 0.0002020655 \\
			& & $0.0014$ & $0.00015$ & $\log(0.5)$ & 0 & $\log(0.6)$ & $\log(0.6)$ & 0 & $\log(3)$ & 0 & 0.0013875856 \\
			
			\midrule
			
			\multirow{4}{*}{\textbf{5}} & \multirow{2}{*}{0} & $0.00014$ & $0.00015$ & $\log(0.445)$ & $\log(0.7)$ & 0 & 0 & $\log(0.7)$ & 0 & 0 & 0.0002053677 \\
			& & $0.0014$ & $0.00015$ & $\log(0.445)$ & $\log(0.7)$ & 0 & 0 & $\log(0.7)$ & 0 & 0 & 0.0014970387 \\
			\cmidrule{2-12}
			& \multirow{2}{*}{$\log(0.2)$} & $0.00014$ & $0.00015$ & $\log(0.445)$ & $\log(0.7)$ & 0 & 0 & $\log(0.7)$ & 0 & 0 & 0.0001892460 \\
			& & $0.0014$ & $0.00015$ & $\log(0.445)$ & $\log(0.7)$ & 0 & 0 & $\log(0.7)$ & 0 & 0 & 0.0012559950 \\
			
			\midrule
			
			\multirow{4}{*}{\textbf{6}} & \multirow{2}{*}{0} & $0.00014$ & $0.00015$ & $\log(0.445)$ & $\log(0.7)$ & 0 & 0 & 0 & 0 & $\log(0.7)$ & 0.0001953243 \\
			& & $0.0014$ & $0.00015$ &  $\log(0.445)$ & $\log(0.7)$ & 0 & 0 & 0 & 0 & $\log(0.7)$ & 0.0014426110 \\
			\cmidrule{2-12}
			& \multirow{2}{*}{$\log(0.2)$} & $0.00014$ & $0.00015$ & $\log(0.445)$ & $\log(0.7)$ & 0 & 0 & 0 & 0 & $\log(0.7)$ & 0.0001825023 \\
			& & $0.0014$ & $0.00015$ & $\log(0.445)$ & $\log(0.7)$ & 0 & 0 & 0 & 0 & $\log(0.7)$ & 0.0011991308 \\
			
			\bottomrule 
		\end{tabular}}
		\medskip
		\caption{An overview of all combinations of parameters used in each scenario of the simulation study, when simulating data for $Y_t$ based on an Aalen additive hazards model. Simulation parameters that were fixed over all scenarios are not included. $\lambda_C$ was determined by first simulating a single dataset with $1,000,000$ individuals from the respective DGP and then using the Newton-Raphson algorithm to numerically determine the rate needed for an exponential distribution that leads to approximately 30\% censoring.}
		\label{tab::dgp_params_aalen}
	\end{table}
	
\end{landscape}

\FloatBarrier

\section{Extensions to Continuous Time}

The results presented in this article are easily generalizable to the setting in which time is continuous instead of discrete. First, we re-define the target estimands as stochastic processes. with $t \in [0, \tau]$. The counterfactual processes for the time-dependent confounders $L_1^{a, a'}, L_2^{a, a'}, ..., L_w^{a, a'}$ are then given as:

\begin{equation}
	\frac{d}{dt} L_{i}^{a,a'}\!(t) = \mathcal{L}_{i}\!\left(t, a, L_i^{a,a'}\!(t), M^{a,a'}\!(t), U_i(t)\right)
\end{equation}

with the counterfactual mediator process $M^{a, a'}(t)$ being defined as:

\begin{equation}
	\frac{d}{dt} M^{a,a'}\!(t) = \mathcal{M}\!\left(t, a', \mathbf{L}^{a,a'}\!(t), M^{a,a'}\!(t), U_M(t)\right),
\end{equation}

and the counterfactual outcome process $Y^{a, a'}(t)$ being defined as:

\begin{equation}
	\frac{d}{dt} Y^{a,a'}\!(t) = \mathcal{Y}\!\left(t, a, \mathbf{L}^{a,a'}\!(t), M^{a,a'}\!(t), Y^{a, a'}\!(t), U_Y(t)\right)
\end{equation}

The initial conditions for $t = 0$ are defined equivalently to the discrete-time case as: $L_i^{a, a'}\!(0) = L_{i,0}, M^{a, a'}\!(0) = M_{0}$ and $Y^{a, a'}\!(0) = Y_{0}$. The hazard function for $T$ (the first time that $Y^{a, a'}(t)$ turns 1) is then defined as:

\begin{equation}
	\lambda_Y^{a,a'}(t)
	=
	\lim_{\Delta t \downarrow 0}
	\frac{
		P\!\left(
		t \le T_Y^{a,a'} < t+\Delta t
		\;\middle|\;
		T_Y^{a,a'} \ge t
		\right)
	}{\Delta t}.
\end{equation}

The target estimands defined in equations~\ref{eq::target_TE}, \ref{eq::target_DE} and \ref{eq::target_IE} remain unchanged, once the hazard function definition has been updated. The identifiability assumptions remain largely equivalent, as shown by \textcite{Kateline2025}. This generalization to continuous time does not change how the Cox, Aalen or AFT model based difference method may be applied, because the employed models were directly designed to handle continuous time. In practice, researchers may structure the data in the counting process format to fit the models. In this format, only one row per time-interval without covariate changes, with associated beginning and end times, is included in the dataset. This is the same format that we used for the discrete-time case.
\par
Applying the parametric g-formula is a little more complicated in continuous time, but not impossible. First, the required models for the conditional probabilities of $Y_t$, $L_{1, t}, L_{2, t}, ..., L_{w, t}$ and $M_t$ may be approximated by discretizing the time into a sufficiently large number of equally sized bins, structuring the data in the long-format and fitting the models as in the discrete-time case. The method may then be applied as outlined in the main text. Alternatively, under the assumptions used throughout the article, the data may also be kept in the counting process format, by fitting weighted models instead, where the weights correspond to the duration of each time-interval in the data. Note that under more realistic assumptions, such as time-varying baseline hazards or complex multilevel structures, the latter strategy may no longer be feasible. After obtaining suitable parametric models, exact realizations of the stochastic processes, as approximated by the models, may then be generated using the discrete-event simulation approach outline in appendix~\ref{appendix::dgp}. The simulated datasets may then be analysed using the same univariable models as in the discrete-time case to obtain the final estimates.
\par
Although we did not present the results here, the simulation study performed in this study is very close to the continuos time case. In fact, since we used a discrete-event simulation algorithm, the only thing that would need to be changed to run the same simulation in continuous time is to remove step 8 in the process used to generate the data (see appendix~\ref{appendix::dgp}).

\end{document}